\pdfoutput=1
\documentclass[aps,prb,floatfix,amsmath,amssymb,preprint,eqsecnum,nofootinbib]{revtex4}
\usepackage{graphicx}
\usepackage{dcolumn}
\usepackage{bm}
\usepackage{setspace}   
\usepackage{epstopdf}
\usepackage{bbm}
\usepackage{amsmath}
\usepackage{amssymb}

\DeclareMathOperator{\re}{Re}
\DeclareMathOperator{\im}{Im}

\usepackage{feynmp}
\newcommand{\mufl}{\mu_{\mathrm{fl}}}
\newcommand{\kq}{\left(\frac{2}{3}\pi, \frac{2}{3}\pi\right)}

\begin{document}
\title{Insulator-metal transition on the triangular lattice}

\author{Yang Qi}
\affiliation{Department of Physics,
Harvard University, Cambridge MA 02138, USA}

\author{Subir Sachdev}
\affiliation{Department of
Physics, Harvard University, Cambridge MA 02138, USA}

\date{\today\\[24pt]}

\begin{abstract}
Mott insulators with a half-filled band of electrons on the triangular lattice
have been recently studied in a variety of organic compounds. All of these
compounds undergo transitions to metallic/superconducting states under
moderate hydrostatic pressure. We describe the Mott insulator using its hypothetical
proximity to a $Z_2$ spin liquid of bosonic spinons. This spin liquid has
quantum phase transitions to descendant
confining states with N\'eel or valence bond solid order, and the insulator
can be on either side of one of these transitions.
We present a theory of fermionic charged excitations in these states, and describe the 
route to metallic states with Fermi surfaces. We argue that an excitonic condensate can
form near this insulator-metal transition, due to the formation of charge neutral
pairs of charge $+e$ and charge $-e$ fermions. This condensate breaks the lattice space
group symmetry, and we propose its onset as an explanation of a low temperature anomaly
in $\kappa$-(ET)$_2$Cu$_2$(CN)$_3$. We also describe the separate BCS instability
of the metallic states to the pairing of like-charge fermions and the onset of superconductivity.
\end{abstract}

\maketitle

\section{Introduction}
\label{sec:intro}

The organic superconductors have proved to be valuable systems for exploring
strong correlation physics near the metal-insulator transition: it is possible to tune
parameters across the metal-insulator transition by moderate pressure, while maintaining
a commensurate density of carriers. Of particular interest are the series of compounds which
realize a half-filled band of electrons on a triangular lattice. In the insulator, three distinct 
ground states are realized in closely related compounds: ({\em i\/}) a N\'eel ordered state \cite{kanoda1} 
in $\kappa$-(ET)$_2$ Cu[N(CN)$_2$]Cl
({\em ii\/}) a recently discovered valence bond solid (VBS) state \cite{kato1,kato2,kato3} in EtMe$_3$P[Pd(dmit)$_2$]$_2$, 
and ({\em iii\/}) an enigmatic
spin liquid state \cite{kanoda0,kanoda2,palee} in $\kappa$-(ET)$_2$Cu$_2$(CN)$_3$. 
Under pressure, all these compounds undergo a transition to a superconducting state. 

The nearest-neighbor $S=1/2$ Heisenberg antiferromagnet on the triangular lattice is believed
to have a small, but non-zero, average magnetic moment \cite{singh,misguich,motrunich,senechal}. However, perturbations such as
ring and second-neighbor exchange are likely to be present near the metal-insulator transition, and these
presumably yield the non-magnetic states noted above. This paper will examine the organic
compounds from the perspective of a particular spin-liquid insulating ground state:
the $Z_2$ spin liquid of 
bosonic spinons proposed in Refs.~\onlinecite{sstri,ashvin}. An advantage of this state is that it 
is a very natural starting point for building the entire phase diagram: theories of the transitions out
of this $Z_2$ spin liquid state into the magnetically ordered (N\'eel) state \cite{css} and the valence
bond solid state \cite{ms} have already been presented, thus realizing the three classes of insulators
which are observed in these compounds. 

The core analysis of this paper concerns the insulator-to-metal transition on the triangular lattice \cite{senechal}
out of the $Z_2$ spin liquid state. The
spin-charge separation of the insulating spin liquid state (and the associated $Z_2$ topological order) will survive
across the transition into the metallic state, and so this is a insulator-metal transition between two `exotic' states.
However, our theory generalizes easily to the other `non-exotic' insulators with conventional order noted above: in these
cases it is the conventional order which survives across the transition, rather than the topological order. Thus our study of the
$Z_2$ spin liquid state offers a convenient context in which to study the insulator-metal transition, with a focus on the charged
excitations in the simplest situation. Subsequently, we can include 
the spin-sector instabilities of the $Z_2$ spin liquid to conventionally ordered confining
states; these do not modify the physics of the insulator-metal transition in a significant manner.

Our perspective on the insulator-metal transition should be contrasted to that in dynamical mean-field theory \cite{dmft,ross} (DMFT).
DMFT begins with a correlated metallic state, and presents a theory of the metal-to-insulator transition. Partly as a consequence
of its focus on the metal, it does a rather poor job of capturing the superexchange interactions in the insulator, and the consequent
resonating valence bond (RVB) correlations. Instead, we begin with a theory of RVB-like correlations in the insulator,
and then develop a theory of the charged excitations; the closing of the gap to these charged excitations describes the
insulator-to-metal transition. A theory of the closing of the charge gap in an RVB insulator has also been discussed recently
by Hermele \cite{hermele}, but for an insulator with gapless fermionic spinons on the honeycomb lattice.

For a half-filled band, our main result is that prior to reaching the metallic state, the insulator has an instability
to a state with {\em exciton condensation}\cite{rice}, associated with the pairing of fermions of charge $e$ 
with fermions of charge $-e$. As in the conventional theory \cite{rice}, the pairing arises because of the attractive
Coulomb interaction between these oppositely charged fermions, while the charge gap to these fermionic excitations 
is still finite, but small.
The excitonic insulator maintains the charge gap, but is associated with
a breaking of the lattice space group symmetry which we shall describe. Eventually, Fermi surfaces of charge $\pm e$
fermions appear, leading to a metallic state in which the excitonic instability (and the lattice symmetry breaking) can also survive. 

We propose that the above instability offers a resolution of a central
puzzle in the properties of the `spin liquid' compound $\kappa$-(ET)$_2$Cu$_2$(CN)$_3$.  
At a temperature around 10K a feature is observed \cite{kanoda2,kanoda3} in nearly all observables suggesting a phase transition 
into a mysterious low temperature correlated state.
An important characteristic of this
compound is the indication from optical conductivity measurements \cite{kanoda4} that the charge gap of this
insulator is quite small, suggesting its proximity to the metal-insulator transition even at ambient
pressure. We propose the 10K transition occurs in the charge sector, and is a consequence of the appearance of the excitonic 
insulator. Observation of the lattice symmetry breaking associated with the excitonic insulator state is an obvious
experimental test of our proposal. There is a suppression of the spin susceptibility below the transition, and this can arise
in our theory from the strong coupling between the fermionic charge carriers and the spinons; this coupling is given by the
large hopping term, $t$ in a $t$-$J$ model, and we will discuss a number of consequences of this term.

The $Z_2$ spin liquid state upon which our excitonic insulator state is based has a gap to all spin excitations,
and this is potentially an issue in applications to $\kappa$-(ET)$_2$Cu$_2$(CN)$_3$.
However, as we have already mentioned above, our theory is easily extend across the critical point to the
onset of magnetic order. We propose that the spin gap is very small due to proximity to this
critical point, and that along with the effect of impurities, this can possibly explain the large
density of states of gapless spin excitations observed in Ref.~\onlinecite{kanoda2}. Indeed, the
$T$ dependence of the spin susceptibility fits the
predictions of the nearest-neighbor Heisenberg antiferromagnet reasonably well, and the latter model
is believed to be very close to a quantum phase transition at which the N\'eel order disappears \cite{singh}. 

The central focus of this paper will be on the charged excitations of the $Z_2$ spin liquid.
These will spinless fermions of charge $\pm e$: we call the charge $+e$
fermion a `holon', and the charge $-e$ fermion a `doublon'. We will begin with a theory of the dispersion
spectrum of single holons and doublons in Section~\ref{sec:holon}. The remaining sections of the paper will consider
various many-body states that can appears in a finite density liquid of these fermions. The states we will find are:\\
({\em i\/}) {\bf Excitonic condensates}: This will be discussed in Section~\ref{sec:mft-exciton}.
Here there is an instability towards the pairing between a holon and doublon to form a condensate
of neutral bosons. The only physical symmetry broken by this pairing is the space group symmetry
of the triangular lattice, and so the state with an excitonic condensate has a spatial modulation
in the electronic charge density. We find this state as an instability of the insulating $Z_2$ spin liquid,
which then reaches an insulating state with an excitonic condensate. However, eventually Fermi surfaces
of holons and doublons can develop, leading to a metallic state with an excitonic condensate.\\
({\em ii\/}) {\bf Metals}: As the repulsion energy $U/t$ in a Hubbard model becomes smaller,
we will find that Fermi surfaces of the holons and doublons can appear (as has just been noted
in the discussion of excitonic condensates). Such states are realizations of `algebraic charge liquids' (ACLs)
introduced recently in Ref.~\onlinecite{acl}, and these will be discussed further
in Section~\ref{sec:lutt}. The simplest such states inherit the topological order of the $Z_2$ spin liquid.
The areas enclosed by the Fermi surfaces in these fractionalized metallic states obey a modified Luttinger
theorem, whose form will be discussed in Section~\ref{sec:lutt}. We will also discuss the possible
binding of holons and doublons to spinons, and the eventual appearance of conventional Fermi liquid
metallic states. \\
({\em iii\/}) {\bf Superconductors}: The metallic states above are also susceptible to 
a conventional BCS instability to a superconducting state, by the condensation of bosons
of charge $\pm 2e$. This will be discussed in Section~\ref{sec:sc}. The required
attractive interactions between pairs of holons (or pairs of doublons) is driven by the exchange of 
spinon excitations of the $Z_2$ spin liquid. We will compute this interaction, and show that it favors
a superconductor in which the physical electronic pair operators have a $d+ i d'$ pairing signature.

\section{Charged excitations of the $Z_2$ spin liquid}
\label{sec:holon}

As we noted in Section~\ref{sec:intro}, the central actor in our analysis, from which
all phases will be derived, is the insulating $Z_2$ spin liquid state on the triangular
lattice \cite{sstri,ashvin}. A review of its basic properties, using the perspective
of the projective symmetry group \cite{wenpsg} is presented in Appendix ~\ref{sec:mfa}. 

Here we will extend the $S=1/2$ antiferromagnet to the 
$t$-$J$ model on the triangular lattice and describe the spectrum of the
charged holon and doublon excitations of the $Z_2$ spin liquid. 
Because of the spin-charge
separation in the parent $Z_2$ spin liquid, both the holons and doublons are spinless, but carry a unit charge
of the $Z_2$ gauge field of the spin liquid. The distinct metallic, insulating, and superconducting
phases which can appear in a system with a finite density of holons and doublons are described
in the subsequent sections.  

We consider the usual $t$-$J$ model on a triangular lattice:
\begin{equation}
  \label{eq:tJ}
  H=-t\sum_{\left<ij\right>,\sigma}c_{i\sigma}^\dagger c_i^\sigma +
  J\sum_{\left<ij\right>}\bm{S}_i\cdot\bm{S}_j
\end{equation}
where $i,j$ are sites of the triangular lattice.
As discussed in Appendix~\ref{sec:mfa}, the $Z_2$ spin liquid is realized by 
expressing the the spins in terms of Schwinger bosons $b_{i\alpha}$.
Here, we introduce spinless, charge $e$, fermionic holons, $f_i$, and 
spinless, charge $-e$, fermionic doublons, $g_i$. The $b_{i\alpha}$, $f_i$, and $g_i$
all carry a unit $Z_2$ charge. At half-filling, the density of holons
must equal the density of doublons, but our formalism also allows consideration 
of the general case away from half-filling with unequal densities.
We can now expression the electron annihilation operator $c_{i \alpha}$ in terms of
the these degress of freedom by
\begin{equation}
  \label{eq:sc-sep}
  c_{i\alpha}^\dagger =f_ib_{i\alpha}^\dagger + g_i^\dagger\epsilon_{\alpha\beta}b_i^\beta
\end{equation}
where $\epsilon_{\alpha\beta}$ is the antisymmetric unit tensor. A similar spin-charge decomposition
of the electron into spinons was discussed in the early study of Zou and Anderson \cite{Zou1988}, but with the opposite
assignment of statistics {\em i.e.\/} fermionic spinons and bosonic holons. In our case, the choice of statistics
is dictated by the choice of the $Z_2$ spin liquid we are considering.

\begin{figure}[htbp]
  \centering
  \includegraphics{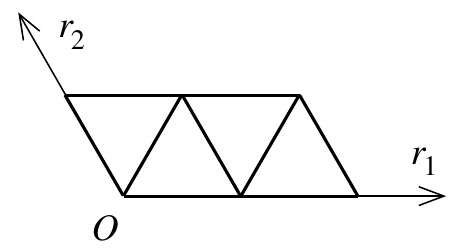}
  \caption{Oblique coordinates system of triangle lattice\cite{ashvin}}
  \label{fig:coord}
\end{figure}

With the spin liquid background, the charge degrees of freedom
acquire a kinetic energy from the hopping term. To see this, plug the
separation (\ref{eq:sc-sep}) into the $t$-term in equation
(\ref{eq:tJ})
\begin{align}
  \nonumber H_t=&-t\sum_{\left<ij\right>}\left(f_ib_{i\alpha}^\dagger +
    \epsilon_{\alpha\beta}g_i^\dagger b_i^\beta\right)\cdot\left(
    f_j^\dagger b_j^\alpha+
    \epsilon^{\alpha\gamma}g_jb_{j\gamma}^\dagger\right)
  +\text{h. c.} \\
  \nonumber
  =&-t\sum_{\left<ij\right>}f_if_j^\dagger b_{i\alpha}^\dagger
  b_j^\alpha
  -t\sum_{\left<ij\right>}\epsilon_{\alpha\beta}\epsilon^{\alpha\gamma}
  g_i^\dagger g_jb_i^\beta b_{j\gamma}^\dagger
  -t\sum_{\left<ij\right>}f_ig_j\epsilon^{\alpha\gamma}
  b_{i\alpha}^\dagger b_{j\gamma}^\dagger \\
  \label{eq:tterm}
  &-t\sum_{\left<ij\right>}g_i^\dagger f_j^\dagger
  \epsilon_{\alpha\beta} b_i^\beta b_j^\alpha
  +\mbox{h. c.}
\end{align}
Replacing the spinon operators with the mean field expectation values
of the spinon pair operators discussed in Appendix~\ref{sec:mfa}, we obtain
\begin{align*}
  H_t^{\text{MF}}=&-t\sum_{\left<ij\right>}f_if_j^\dagger B_{ij}
  -t\sum_{\left<ij\right>}g_i^\dagger g_jB_{ji}
  -t\sum_{\left<ij\right>}f_ig_jA_{ij}^\ast
  -t\sum_{\left<ij\right>}g_i^\dagger f_j^\dagger A_{ji}
  +\text{h. c.} \\
  =&-Bt\sum_{\left<ij\right>}\left(f_if_j^\dagger+f_jf_i^\dagger\right)
  -Bt\sum_{\left<ij\right>}\left(g_i^\dagger g_j+g_j^\dagger
    g_i\right) \\
  &-t\sum_{\left<ij\right>}A_{ij}\left(g_j^\dagger
    f_i^\dagger-g_i^\dagger f_j^\dagger\right)
  -t\sum_{\left<ij\right>}A_{ij}^\ast\left(f_ig_j-f_jg_i\right)
\end{align*}
where $A_{ij}$ and $B_{ij}$ are the expectation values defined in Eq.~(\ref{eq:ansatz}).
Using Fourier transform to convert this to momentum space, we obtain
\begin{equation}
  \label{eq:htmf}
  H_t^{\text{MF}}=-2tB\sum_k\left(f_kf_k^\dagger+g_k^\dagger
    g_k\right)\re{\xi_k}
  +2tA\sum_k\left(g_k^\dagger f_{-k}^\dagger+f_{-k}g_k\right)\im\xi_k
\end{equation}
where $\xi_k$ is defined as
\begin{equation}
  \label{eq:xik}
  \xi_k=e^{ik_1}+e^{ik_2}+e^{ik_3}
\end{equation}
Here we are using the oblique co-ordinate
system shown in
Fig. \ref{fig:coord}. For momentum space, we use the dual base with
respect to the coordinate system. Therefore
$\bm{k}\cdot\bm{r}=k_1r_1+k_2r_2$. Momenta $(k_1,k_2)$,
$(k_1+2\pi,k_2)$ and $(k_1, k_2+2\pi)$ are equivalent. For convenience,
a third component of momentum is defined as $
  k_3=-k_1-k_2$
to make formulas more symmetric.

In the complete Hamiltonian, we need a chemical potential $\mu$ to couple
to the total charge density: when this is within the Mott gap, the density of holons
and doublons will be equal and the system will be half-filled.
Also, the double occupancy
energy $U$ needs to be included since the number of double occupied
sites is not conserved. Combining these two terms with the kinetic
energy (\ref{eq:htmf}), we obtain
\begin{equation}
  \label{eq:hmf}
  \begin{split}
    H^{\text{MF}}=&-2tB\sum_k\left(f_kf_k^\dagger+g_k^\dagger
      g_k\right)\re{\xi_k}
    +2tA\sum_k\left(g_k^\dagger
      f_{-k}^\dagger+f_{-k}g_k\right)\im\xi_k \\
    &+\mu\sum_k\left(g_k^\dagger g_k-f_k^\dagger
      f_k\right)+U\sum_kg_k^\dagger g_k
  \end{split}
\end{equation}

The above Hamiltonian can be diagonalized by Bogoliubov
transformation. To do so, first write the Hamiltonian in the following
matrix form:
\begin{equation}
  \label{eq:hmf-matrix}
  H^{\text{MF}}=\sum_k\Psi_k^\dagger D_k\Psi_k
\end{equation}
where
\begin{equation}
  \label{eq:psik}
  \psi_k=\left(g_k, f_{-k}^\dagger\right)^T
\end{equation}
and the kinetic energy matrix
\begin{equation}
  \label{eq:Dk}
  D_k=\left(\begin{matrix}
      -2tB\re\xi_k+\mu+U & 2tA\im\xi_k\\
      2tA\im\xi_k & -2tB\re\xi_k+\mu
    \end{matrix}\right)
\end{equation}
The spectrum of quasi-particles is the eigenvalue of the matrix
(\ref{eq:Dk})
\begin{equation}
  \label{eq:spectrum}
  E_k=-2tB\re\xi_k+\mu+\frac{U}{2}\pm\sqrt{\frac{U^2}{4}-4t^2A^2(\im\xi_k)^2}
\end{equation}

We can redefine the chemical potential $\mu$ to absorb the
$\frac{U}{2}$ term in Eq.~(\ref{eq:spectrum}). Also, to make it
simpler, define
\begin{equation}
  \label{eq:mufl}
  \mufl=\frac{U}{2}
\end{equation}
Then the dispersion of the two bands looks like
\begin{equation}
  \label{eq:mix}
  E_k^\pm=-2tB\re\xi_k\pm\sqrt{\mufl^2+(2tA\im\xi_k)^2}
\end{equation}
To study the maximum and minimum position of the about dispersion, we
need to first identify the stationary points of the dispersion in the
Brillouin zone. This can be done by noticing that the two $k$
dependent function $\re\xi_k$ and $\im\xi_k$ in equation
(\ref{eq:mix}) reach stationary points simultaneously at the points
listed in the first column of Table~\ref{tab:stpoints}.
\begin{table}[htbp]
  \centering
  \begin{tabular}{c|c|c|c}
    \hline
    Steady points & $\re\xi_k$ & $|\im\xi_k|$ & $E_k^\pm$ \\
    \hline
    $\pm\left(\frac{2}{3}\pi,\frac{2}{3}\pi\right)$ & $-\frac{3}{2}$ &
    $\frac{3}{2}\sqrt{3}$ & 
    $3tB\pm\sqrt{\mufl^2+\left(3\sqrt{3}tA\right)^2}$ \\
    $(0,0)$ & 3 & 0 & $-6tB\pm\mufl$\\
    $(\pi,\pi)$, $(0,\pi)$, $(\pi,0)$ & -1 & 0 & $2tB\pm\mufl$\\
    \hline
  \end{tabular}
  \caption{Stationary points of $\xi_k$.}
  \label{tab:stpoints}
\end{table}
The maximal and minimal points of the bands must come from the
candidates listed in Table~\ref{tab:stpoints}. To determine which one
is the maximum or minimum, we compare the energy at those stationary
points, which is listed in Table~\ref{tab:stpoints} along with real
and imaginary parts of $\xi_k$.
It is easy to see that the position of minimum and maximum depends on
the sign of $tB$ and the value of $\mufl$, $tA$ and $tB$. Let us discuss
two extreme case $\mufl\ll tA, tB$ and $\mufl\gg tA, tB$ for both $tB>0$ and
$tB<0$.

\begin{itemize}
\item $tB>0$

First, consider the extreme case $\mufl\gg tA$. In this case the
effect of the $tA$ term can be ignored comparing to the $\mufl$ term
in the square root and the shape of dispersion is solely determined by
the $tB\re\xi_k$ term, which is listed in the second column of table
\ref{tab:stpoints}. Therefore the maximums locate at $\pm\kq$, and the
minimal point is $(0,0)$, for both the upper and lower bands.

Then consider the other extreme: $\mufl=0$. In this case the energy at
the stationary points are
\begin{align}
  \label{eq:epm2}
  E_k^\pm=3tB\pm3\sqrt{3}|tA|,&\quad k=\pm\kq \\
  \label{eq:epm0}
  E_k^\pm=-6tB,&\quad k=(0,0) \\
  \label{eq:epm3}
  E_k^\pm=2tB,&\quad k=(\pi,\pi),(0,\pi),(\pi,0)
\end{align}
Without the $tA$ term, (\ref{eq:epm2}) was the maximum. However, when
the $3\sqrt{3}tA$ term is deducted from it, it becomes smaller and can
be smaller than (\ref{eq:epm3}). Then the maximal points of the lower
band become $(\pi,\pi)$, $(0,\pi)$ and $(\pi,0)$. The requirement for
(\ref{eq:epm3}) to be smaller than (\ref{eq:epm2}) is
\begin{equation}
  \label{eq:requireAB}
  |B|<\sqrt{3}|A|
\end{equation}
By minimizing the Hamiltonian the mean field parameters $A$ and $B$ we
get show that the inequality (\ref{eq:requireAB}) holds. Therefore the
minimal points are $(\pi,\pi)$, $(0,\pi)$ and $(\pi,0)$.

On the other hand, for the upper band the $tA$ term makes equation
(\ref{eq:epm2}) larger so that it will never be smaller than
(\ref{eq:epm0}). Consequently the minimal point of upper band remains
at $(0,0)$. These results are summarized in Table~\ref{tab:minmax}.

\item $tB<0$

For this case, notice that the dispersions of the bands for parameters
$tB>0$ and $tB<0$ are related
\begin{equation}
  \label{eq:relate}
  E_k^\pm(tB)=-E_k^\mp(-tB)
\end{equation}
Therefore for $B<0$ the bands are up-side down. Hence the bottom of
upper band and the top of the lower band are switched (see Table~\ref{tab:minmax}).
\end{itemize}

\begin{table}[htbp]
  \centering
  \begin{tabular}{c|c|c|c}
    \hline
    $tB$ & $\mufl$ & Min of $\epsilon^+$ & Max of $\epsilon^-$ \\
    \hline
    $tB>0$ & $\mufl\gg |tA|$ & $(0,0)$ & $\pm\kq$ \\
    $tB>0$ & $\mufl\ll |tA|$ & $(0,0)$ & $(0,0),(0,\pi),(\pi,0)$ \\
    \hline
    $tB<0$ & $\mufl\gg |tA|$ & $\pm\kq$ & $(0,0)$ \\
    $tB<0$ & $\mufl\ll |tA|$ & $(0,0),(0,\pi),(\pi,0)$ & $(0,0)$ \\
    \hline
  \end{tabular}
  \caption{Minimal and maximal points of the bands}
  \label{tab:minmax}
\end{table}

\begin{figure}[htbp]
  \centering
  \includegraphics{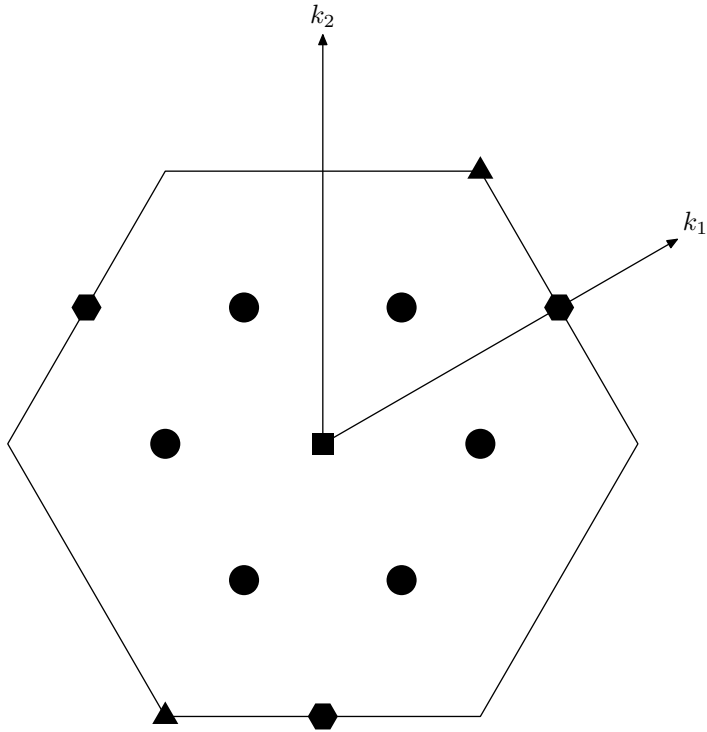}
  \caption{Brillouin zone of the triangular lattice. The first
    Brillouin zone has the hexagon shape, and the two axes are the
    dual base with respect to the oblique coordinates we are using in
    Fig.~\ref{fig:coord}. The symbols corresponds to the momenta of
    minimal energy of excitations. The solid triangles at
    $\pm\left(\frac{2\pi}{3}, \frac{2\pi}{3}\right)$ are spinon
    excitations (see Appendix~\ref{sec:mfa}); the solid hexagons at
    $(\pi,0)$, $(0, \pi)$ and $(\pi, \pi)$ are doublons $g$; the solid
    square at the origin is the holon $f$. The hole (bound state of holon and
    spinon) is located at the same position as the spinon (triangles);
    the electron (bound state of doublon and spinon) locates at the
    solid circles: $\pm\left(-\frac{\pi}{3}, \frac{2\pi}{3}\right)$,
    $\pm\left(\frac{2\pi}{3}, -\frac{\pi}{3}\right)$ and
    $\pm\left(\frac{\pi}{3}, \frac{\pi}{3}\right)$. The bound states
    are discussed in Appendix \ref{sec:boundstate}.}
  \label{fig:hexagon}
\end{figure}

According to Eq.~(\ref{eq:relate}), the dispersion of the bands
are flipped when $tB$ changes the sign. However, since we are at zero
doping, the flip of the bands only switches the name of holons and
doublons, but the physics is very similar in the two cases. From
minimizing the spinon mean field Hamiltonian we found that
$B<0$. Therefore in the following discussion we will assume that $B<0$
and $t>0$. Consequently the maximum of valence band and the minimum of
electron band comes from the $tB<0$ row of Table~\ref{tab:minmax}:
when the band gap is small $\mufl\ll|tA|$, lower band excitations are
around $(0,0)$, and higher band excitations are around $(0,\pi)$,
$(\pi,0)$ and $(\pi,\pi)$. It is worth noticing that the mixing term
in Eq.~(\ref{eq:hmf}), which is proportional to $tA\im\xi_k$,
vanishes at these points (see Table~\ref{tab:stpoints}). Therefore, the
Bogoliubov transformation is trivial at the center of the Fermi sea of the two types of
excitations. If we assume that the Fermi sea of excitations are small,
then the excitations can be identified as holons and
doublons. According to the definition of quasiparticle
(\ref{eq:psik}), it carries the same electric charge as the original
electron. Therefore the quasiparticle excitation in the upper band is
doublon, and the quasihole excitation in the lower band is
holon. Therefore low energy effective fields of holon ($f$) and doublon ($h$) can
be defined as
\begin{align}
  \label{eq:field-holon}
  f(\bm{r})&=\sum_qf_qe^{i\bm{q}\cdot\bm{r}} \\
  \label{eq:field-doublon}
  g_\lambda(\bm{r})&=\sum_qg_{k_\lambda^D+q}e^{i\bm{q}\cdot\bm{r}},\quad \lambda=1,2,3
\end{align}
where $\bm{k}_\lambda^D$ denotes three minimums in doublon dispersion:
\begin{equation}
  \label{eq:klambda}
  k_1^D=(0,\pi),\quad k_2^D=(\pi,0),\quad k_3^D=(\pi,\pi)
\end{equation}
The positions of these excitations in the Brillouin zone are indicated in Fig.~\ref{fig:hexagon}.

\subsection{Two-band model}
\label{sec:two-band}

In this section, we setup a two-band model describing holon and
doublon excitations, and the Coulomb interaction between them. In the
next section, this model will be used to investigate 
exciton condensation in the system when the charge gap becomes small.

When the density of charge excitations is small, the Fermi pockets of
spinon and holon are very small in the Brillouin zone. Therefore we
can make the approximation that the dispersion relation in equation
(\ref{eq:mix}) takes a quadratic form in the pocket. Without
losing any generality, we can assume that $t>0$ and $B<0$. So from
Table \ref{tab:minmax} holon pocket is at $(0,0)$, and doublon pockets
are located at $(\pi,0)$, $(0, \pi)$ and $(\pi, \pi)$. Expand the
dispersion relation in Eq.~(\ref{eq:mix}) to the quadratic
order, we have
\begin{equation}
  \label{eq:hband}
  E_k\simeq 6t|B|-\mufl-\frac{3}{2}t|B|(k_x^2+k_y^2)
\end{equation}
near the minimum of holon, and
\begin{equation}
  \label{eq:dband}
  E_k\simeq-2t|B|+\mufl+\left(\frac{8t^2A^2}{\mufl}-\frac{t|B|}{2}\right)
  k_x^2+\frac{3}{2}t|B|k_y^2
\end{equation}
near the minimum $(0, \pi)$ of doublon. The dispersion relation near
the other two doublon minimums can be obtained by simply rotating
Eq.~(\ref{eq:dband}) by $120^\circ$ degrees.

From Eq.~(\ref{eq:hband}) and (\ref{eq:dband}) the gap between
the two bands can be calculated by subtracting the maximum of holon
band from the minimum of doublon band
\begin{equation}
  \label{eq:uind}
  U_{\text{ind}}=2\mufl-8t|B|
\end{equation}
and it is an indirect gap because the maximum of holon band is at $(0,
0)$, while the minimums of doublon bands are at other places: $(0,
\pi)$, $(\pi, 0)$ and $(\pi, \pi)$. If we do not include any
interactions, the system consists two bands: holon band and doublon
band. If the gap $U_{\text{ind}}$ is positive, it is a semiconductor
with an indirect gap. If the gap is negative, it is a semimetal.

From Eq.~(\ref{eq:hband}), we see that the dispersion relation of
holon is isotropic with the effective mass:
\begin{equation}
  \label{eq:mh}
  m_h=\frac{1}{3t|B|}
\end{equation}

From Eq.~(\ref{eq:dband}) we see that the dispersion of doublon
is not isotropic. For the pocket at $(\pi, 0)$, the effective masses in
$x$ and $y$ directions are
\begin{gather}
  \label{eq:mdx}
  m_{dx}=\left(\frac{16t^2A^2}{\mufl}-t|B|\right)^{-1}\\
  \label{eq:eq:mdy}
  m_{dy}=\frac{1}{3t|B|}
\end{gather}
Note, that $m_{dy}=m_h$. When $U_{\text{ind}}$ is tuned around zero,
the Coulomb interaction will lead to exciton instability. The change
of $U_{\text{ind}}$ is small comparing to $t$, so we can assume the
masses are constant in our problem. At the point $U_{\text{ind}}=0$,
we have $\mufl=4t|B|$. Therefore
\begin{equation}
  \label{eq:mass-ratio}
  \frac{m_h}{m_{dx}}=\frac{m_{dy}}{m_{dx}}=
  \frac{4}{3}\left|\frac{A}{B}\right|^2-\frac{1}{3}
\end{equation}
In the spin liquid phase we have $A>|B|$, therefore the above ratio is
larger than 1.

The dispersion relation for the other two doublon pockets can be
obtaned by rotating the pocket by $\pm120^\circ$.

The exciton problem has been well studied in the context of semimetal
to semiconductor transition\cite{rice}, where people usually
use electron operators for both the valence and conducting
band. Therefore here we want to choose our notations to be consistent
with them. To achieve that, we just need to switch the creation and
annihilation operator for holon. To be clear, we use $h$ and $d$ for
new holon and doublon operators respectively.
\begin{equation}
  \label{eq:h_k}
  h_k=f_k^\dagger,\quad d_k=g_k
\end{equation}
For the doublon bands, we measure the momentum from the minimum in the
pockets:
\begin{equation}
  \label{eq:g_k_123}
  d_{1k}=d_{k+(\pi, 0)},\quad d_{2k}=d_{k+(0, \pi)},\quad
  d_{3k}=d_{k+(\pi, \pi)}
\end{equation}
With the help of those effective masses, the dispersion relation for
holon and doublon can be expressed.
\begin{gather}
  \label{eq:epsilon_h}
  \epsilon_h(\bm{k})=U_{\text{ind}}-\frac{k^2}{2m_h}\\
  \label{eq:epsilon_d1}
  \epsilon_{d2}(\bm{k})=\frac{k_x^2}{2m_{dx}}+\frac{k_y^2}{2m_{dy}}\\
  \label{eq:epsilon_d2}
  \epsilon_{d1}(\bm{k})=\frac{(k_x\cos120^\circ-k_y\sin120^\circ)^2}{2m_{dx}}
  +\frac{(k_x\sin120^\circ+k_y\cos120^\circ)^2}{2m_{dy}}\\
  \label{eq:epsilon_d3}
  \epsilon_{d3}(\bm{k})=\frac{(k_x\cos120^\circ+k_y\sin120^\circ)^2}{2m_{dx}}
  +\frac{(-k_x\sin120^\circ+k_y\cos120^\circ)^2}{2m_{dy}}
\end{gather}
Therefore the kinetic energy of the system is
\begin{equation}
  \label{eq:hkin}
  H_{\text{kin}}=\sum_k\epsilon_h(\bm{k})h_k^\dagger h_k
  +\sum_{k,\lambda}\epsilon_{d\lambda}(\bm{k})d_{\lambda k}^\dagger d_{\lambda k}
\end{equation}

The exciton instability is because of the Coulomb interaction between
doublon and holon. In general, the Coulomb interaction is
\begin{equation}
  \label{eq:coul-gen}
  H_{\text{int}}=\frac{1}{2}\sum_{i\neq j,\alpha\beta}\frac{e^2}{|\bm{r}_i-\bm{r}_j|}
  c_{i\alpha}^\dagger c_{i\alpha}c_{j\beta}^\dagger c_{j\beta}
\end{equation}
where the electron density can be expressed by holon and doublon
operators as the following
\begin{equation}
  \label{eq:ne-hd}
  \sum_\alpha c_{i\alpha}^\dagger c_{i\alpha}=h_i^\dagger
  h_i+d_i^\dagger d_i
\end{equation}
Therefore Eq.~(\ref{eq:coul-gen}) becomes
\begin{align}
  \label{eq:coul-gen-hd}
  H_{\text{int}}&=\frac{1}{2}\sum_{i\neq j}\frac{e^2}{|\bm{r}_i-\bm{r}_j|}
  \left(h_i^\dagger h_i+d_i^\dagger d_i\right)\cdot
  \left(h_j^\dagger h_j+d_j^\dagger d_j\right)\\
  \label{eq:coul-gen-hd-q}
  &=\sum_{kk^\prime q}\frac{4\pi e^2}{q^2}
  \left(h_k^\dagger h_{k+q}+d_k^\dagger d_{k+q}\right)\cdot
  \left(h_{k^\prime}^\dagger h_{k^\prime-q}+d_{k^\prime}^\dagger d_{k^\prime-q}\right)
\end{align}

To analyze this interaction, we can use the screened Hartree-Fock
approximation\cite{rice}: For the Hartree term, the unscreened
Coulomb interaction is used; for the exchange term, the screening from
other electrons is included. This approximation results in the
following functional of interaction energy
\begin{gather}
  \label{eq:dir}
  H_{\text{dir}}=\sum_{k,k^\prime,q}\frac{4\pi e^2}{q^2}
  \left(\left<h_k^\dagger h_{k+q}\right>+
    \left<d_k^\dagger d_{k+q}\right>\right)\cdot
  \left(\left<h_{k^\prime}^\dagger h_{k^\prime-q}\right>+
    \left<d_{k^\prime}^\dagger d_{k^\prime-q}\right>\right)\\
  \nonumber
  H_{\text{ex}}=-\sum_{k,k^\prime,q,\alpha\beta}\frac{4\pi e^2}{\epsilon(q)q^2}
  \left(
    \left<h_k^\dagger h_{k^\prime-q}\right>
    \left<h_{k^\prime}^\dagger h_{k+q}\right>
    +\left<d_k^\dagger d_{k^\prime-q}\right>
    \left<d_{k^\prime}^\dagger d_{k+q}\right>\right.\\
  \label{eq:ex}
    \left.+\left<h_k^\dagger d_{k^\prime-q}\right>
    \left<d_{k^\prime}^\dagger h_{k+q}\right>
    +\left<d_k^\dagger h_{k^\prime-q}\right>
    \left<h_{k^\prime}^\dagger d_{k+q}\right>
  \right)
\end{gather}
where $\left<\cdot\right>$ denotes the average under the mean field
state.

The exciton is formed by pairing of holon and doublon. In the interactions
in Eq.~(\ref{eq:dir}) and (\ref{eq:ex}), only the last two terms
in exchange interaction are relevant for holon and doublon
pairing. Therefore in the exciton problem we can include only these
terms and neglect all the others. This is called dominant term
approximation\cite{rice}.
\begin{equation}
  \label{eq:int-dta-ori}
  H_{\text{DTA}}=-\sum_{kk^\prime q}\frac{4\pi
    e^2}{\epsilon(q)q^2}\left(
    \left<h_k^\dagger d_{k^\prime-q}\right>
    \left<d_{k^\prime}^\dagger h_{k+q}\right>
    +\left<d_k^\dagger h_{k^\prime-q}\right>
    \left<h_{k^\prime}^\dagger d_{k+q}\right>
  \right)
\end{equation}
Since the interaction is invariant in parity transformation, equation
(\ref{eq:int-dta}) can be further simplified as
\begin{equation}
  \label{eq:int-dta}
  H_{\text{DTA}}=-2\sum_{kk^\prime q}\frac{4\pi
    e^2}{\epsilon(q)q^2}
  \left<h_k^\dagger d_{k^\prime-q}\right>
  \left<d_{k^\prime}^\dagger h_{k+q}\right>
\end{equation}

The interaction can be further expanded near the minimums of holon and
doublon pockets as the following
\begin{equation}
  \label{eq:int-dta-hd}
  H_{\text{DTA}}=-\sum_{kk^\prime q,\lambda}\frac{8\pi
    e^2}{\epsilon(\bm{q}+\bm{k}^D_\lambda)(\bm{q}+\bm{k}^D_\lambda)^2}
  \left<h_k^\dagger d_{\lambda k^\prime-q}\right>
  \left<d_{\lambda k^\prime}^\dagger h_{k+q}\right>  
\end{equation}
In this equation $k$, $k^\prime$ and $q$ are both small momenta. The
$q$ dependence in the coefficient of this interaction can therefore be
neglected:
\begin{equation}
  \label{eq:ignore-q}
  \frac{8\pi e^2}{\epsilon(\bm{q}+\bm{k}^D_\lambda)
    (\bm{q}+\bm{k}^D_\lambda)^2}\simeq
  \frac{8\pi e^2}{\epsilon(k^D_\lambda)(k^D_\lambda)^2}
\end{equation}
and we define this coefficient as $g$. The interaction in equation
(\ref{eq:int-dta-hd}) can then be expressed as
\begin{equation}
  \label{eq:int-dta-g}
  H_{\text{DTA}}=-g\sum_{kk^\prime q,\lambda}
  \left<h_k^\dagger d_{\lambda k^\prime-q}\right>
  \left<d_{\lambda k^\prime}^\dagger h_{k+q}\right>  
\end{equation}

\section{Exciton condensation}
\label{sec:mft-exciton}

In the previous section we discussed the dispersion relations of holon
and doublon excitations and the interaction between them. If momentum
is measured from the center of the pockets, the holon pocket is a
sphere centered at $(0,0)$, while the doublon pockets are three
ellipses centered at $(0,0)$, long axes of which are $120^\circ$ to
each other. Since the system is neutral, the area of the holon pocket
is three times larger than the area of each doublon pocket. However,
the long axis of doublon pockets can still be longer than the radius
of holon pocket if the aspect ratio of doublon pockets is greater than
3:1. Therefore there can still be pairing at the places where the two
Fermi surfaces are close. Between holon and one type of doublon the
pairing is only good in one direction, but if the holon is paired with
both three types of doublon, there is a good pairing in three
directions, which covers a large portion of the Fermi surface. We will
see that this mismatch in Fermi surfaces lead to a richer phase
diagram compared to the normal-superconducting problem.

Since all the pockets are centered at $(0,0)$ (when momentum is
measured from the minimums of pockets), we assume that the exciton
pairs also have total momentum $(0,0)$. In other words, it is not a
LLFO state in which smaller pocket is shifted toward the edge of
larger pocket. Therefore we have three pairing order parameters
between holon and each type of doublon
\begin{equation}
  \label{eq:delta-123}
  \Delta_\lambda=-g\sum_k\left<h_k^\dagger d_{\lambda k}\right>
\end{equation}
Note, that the pairing is $s$-wave because the interaction
(\ref{eq:int-dta-g}) only has $s$-wave component.

According to this order parameter, the mean field decomposition of the
interaction (\ref{eq:int-dta-g}) is
\begin{equation}
  \label{eq:h-dta-mft}
  H_{\text{DTA}}=\sum_\lambda\Delta_\lambda
  \sum_kd_{\lambda k}^\dagger h_k+\text{h. c.}+
  \frac{1}{g}\sum_\lambda|\Delta_\lambda|^2
\end{equation}

Combining this with the kinetic energy in Eq.~(\ref{eq:hkin}) and
the chemical potential, the full Hamiltonian is
\begin{align}
  \nonumber
  H=&H_{\text{kin}}+H_{\text{DTA}}-\mu\left[\sum_k\left(h_k^\dagger h_k
      +\sum_\lambda d_{\lambda k}^\dagger d_{\lambda k}\right)-N_0\right]\\
  \nonumber
  =&\sum_k[\epsilon_h(\bm{k})-\mu]h_k^\dagger h_k
  +\sum_{k,\lambda}[\epsilon_{d\lambda}(\bm{k})-\mu]d_{\lambda{}k}^\dagger
  d_{\lambda k}\\
  \label{eq:h-mft}
  &+\sum_\lambda\Delta_\lambda
  \sum_kd_{\lambda k}^\dagger h_k+\text{h. c.}+
  \frac{1}{g}\sum_\lambda|\Delta_\lambda|^2+\mu N_0
\end{align}
This Hamiltonian can be diagonalized using the Bogoliubov
transformation. First, write Eq.~(\ref{eq:h-mft}) in the matrix
form:
\begin{equation}
  \label{eq:h-mft-mat}
  H=\sum_k \Psi_k^\dagger M_k \Psi_k+
  \frac{1}{g}\sum_\lambda|\Delta_\lambda|^2+\mu N_0
\end{equation}
where
\begin{equation}
  \label{eq:psi-k}
  \Psi_k=(\begin{matrix}
    h_k & d_{1k} & d_{2k} & d_{3k}
  \end{matrix})^T
\end{equation}
\begin{equation}
  \label{eq:M-k}
  M_k=\left(\begin{matrix}
      \epsilon_h(\bm{k})-\mu & \Delta_1^\ast & \Delta_2^\ast &
      \Delta_3^\ast\\
      \Delta_1 & \epsilon_{d1}(\bm{k})-\mu & 0 & 0 \\
      \Delta_2 & 0 & \epsilon_{d2}(\bm{k})-\mu & 0 \\
      \Delta_3 & 0 & 0 & \epsilon_{d3}(\bm{k})-\mu \\
    \end{matrix}\right)
\end{equation}

A Bogoliubov transformation is then a unitary transformation
\begin{equation}
  \label{eq:U-k}
  \tilde{\Psi}_k=U_k\Psi_k
\end{equation}
that diagonalizes the matrix $M_k$. With the new quasiparticle
operators $\tilde{\Psi}_k$, the Hamiltonian is diagonalized as
\begin{equation}
  \label{eq:h-mft-diag}
  H=\sum_k\sum_{i=1}^4E_{ik}\alpha_{ik}^\dagger \alpha_{ik}+
  \frac{1}{g}\sum_\lambda|\Delta_\lambda|^2+\mu N_0
\end{equation}
where $\alpha_{ik}$ are the four components of $\tilde{\Psi}_k$, and
$E_{ik}$ are the four eigenvalues of the matrix $M_k$. Therefore in
the ground state, the state of quasiparticle $\alpha_{ik}$ is filled
if $E_{ik}<0$, or empty if $E_{ik}>0$. The ground state energy is
\begin{equation}
  \label{eq:hg}
  \left<H\right>=\sum_{E_{ik}<0}E_{ik}+
  \frac{1}{g}\sum_\lambda|\Delta_\lambda|^2+\mu N_0
\end{equation}

Becuase the three doublon pockets are identical, we assume that the
absolute values of the order parameters are the same for three
pockets. Therefore the pairing order parameters can be characterized
by one absolute value and three phases.
\begin{equation}
  \label{eq:delta-theta}
  \Delta_\lambda=\Delta e^{i\theta_\lambda}
\end{equation}

After we have made the approximation that only excitations near the
minimums of the pockets are considered, the kinetic energy
(\ref{eq:hkin}) and interaction energy (\ref{eq:int-dta-g}) has the
$U(1)\times U(1)\times U(1)$ symmetry for changing the phases of three
doublons independently. Therefore the mean field energy
(\ref{eq:h-mft}) does not depend on the phases of order parameters
$\theta_\lambda$. In other words, for this simplified Hamiltonian, a
state with order parameter in Eq.~(\ref{eq:delta-123}) breaks the
$U(1)\times U(1)\times U(1)$ symmetry spontaneously, and the states
with different phases are degenerate. We will see in the next section
that this degeneracy is lifted once we add the other parts of Coulomb
interaction back.

With this $U(1)\times U(1)\times U(1)$ degeneracy, the ground state
energy in Eq.~(\ref{eq:hg}) only depends on the absolute value of
$\Delta$. The ground state is obtained by minimizing the ground state
energy with respect to order parameter $\Delta$. The results are described below

\begin{figure}[htbp]
  \centering
  \includegraphics{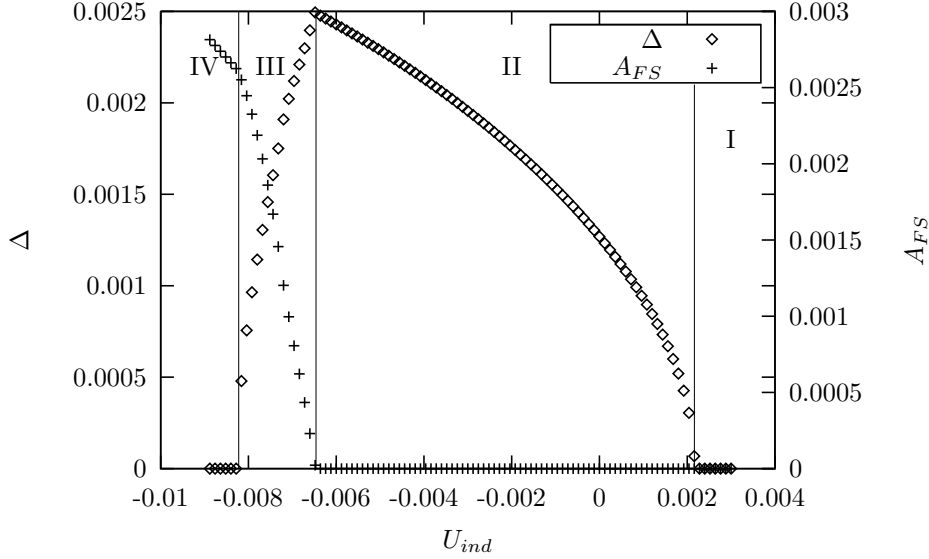}
  \caption{Phase diagram at fixed masses and interaction $g$. This
    plot shows the exciton pairing gap $\Delta$ and the area of
    hole-like quasiparticle Fermi surface as a function of the charge
    gap $U_{\text{ind}}$. It is plotted at $m_h=m_{dy}=3m_{dx}$, and
    $g/m_h=0.007$. $U_{\text{ind}}$ and $\Delta$ are
    measured in unit of the energy scale in the system:
    $\frac{\Lambda^2}{2m_h}$, where $\Lambda$ is the cutoff momentum;
    the Fermi surface area is in unit of $\pi\Lambda^2$. Phase I, II,
    III and IV in the diagram are normal insulating phase, excitonic
    insulator, excitonic metal and normal conducting
    phase respectively.}
  \label{fig:a3scan}
\end{figure}

\begin{figure}[htbp]
  \centering
  \includegraphics{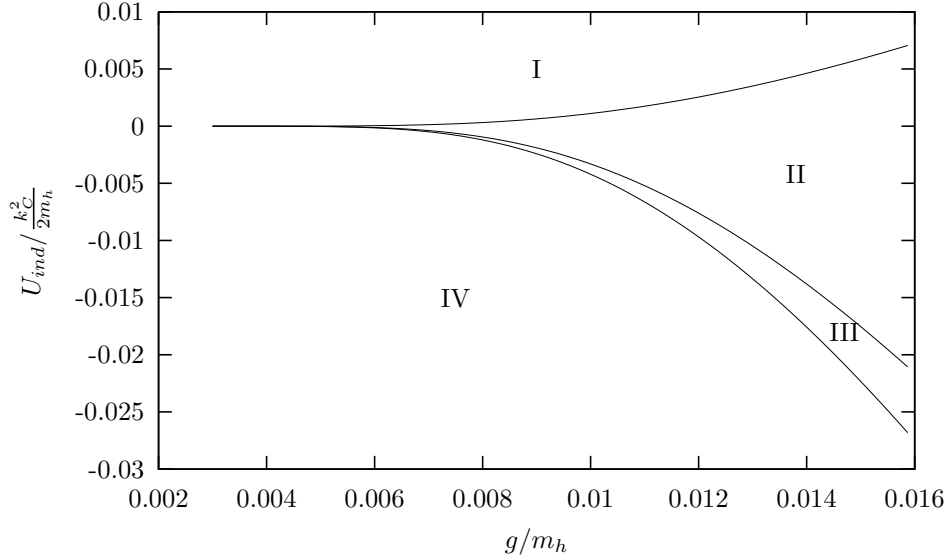}
  \caption{Phase diagram at fixed masses. It is plotted at
    $m_h=m_{dy}=3m_{dx}$. Phase I, II,
    III and IV in the diagram are normal insulating phase, excitonic
    insulator, excitonic metal, and normal conducting
    phase respectively.}
  \label{fig:a3g}
\end{figure}

\begin{figure}[htbp]
  \centering
  \includegraphics{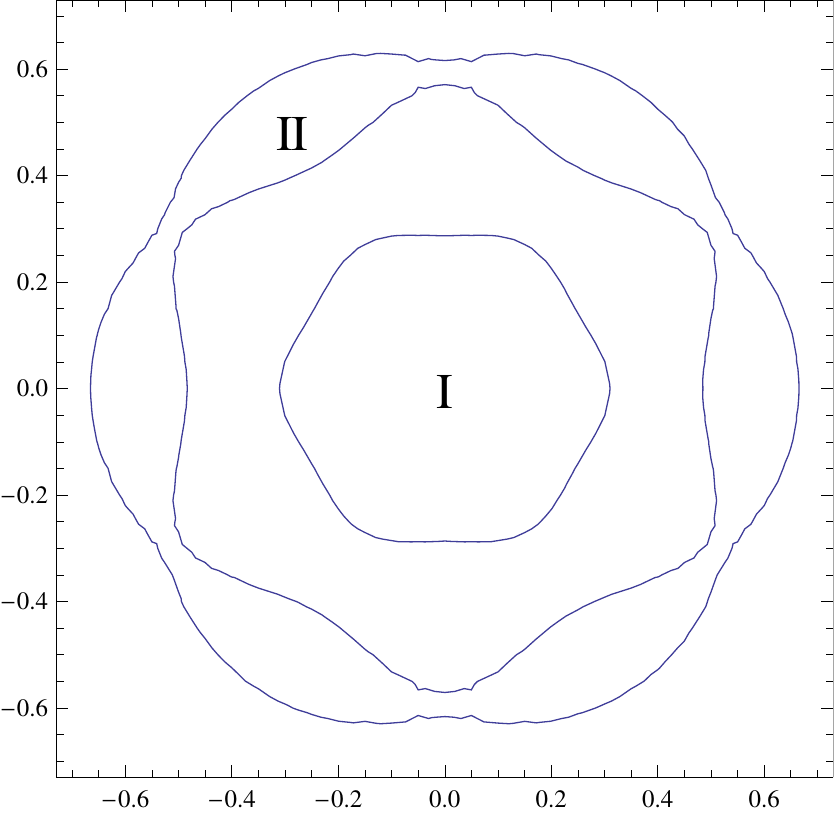}
  \caption{Fermi surface in the excitonic metal. Region I is the
    hole-like Fermi surface of the quasiparticle, and region II is the
    particle-like Fermi surface of the quasiparticle. Because the
    system is charge neutral, the area of region I is equal to the
    area of region II. Note, that this diagram is drawn in the reduced
    Brillouin zone. In presence of the pairing, the holon and doublon
    modes are mixed, and the Brillouin zone shrinks to 1/4 of its
    original size such that the holon and doublon pockets locate at
    the same momenta.}
  \label{fig:exotic-fs}
\end{figure}

When $U_{\text{ind}}$ is large and positive, the system is in an
insulating phase (semiconductor phase), with no Fermi surface and
$\Delta=0$. Then as the gap decreases, the system goes through a
second-order phase transition at a critical value $U_{\text{ind}}^0$
and enters a insulating phase with no Fermi surface but
$\Delta\neq0$. In this phase the binding energy of exciton overcomes
the charge gap, so there are doublon and holon excitations. However,
the excitations near the Fermi surface bind into exciton, and the gap
is large comparing to the Fermi energy such that the Fermi surface is
fully gapped although there is a mismatch between holon and doublon
Fermi surfaces. This state is an insulating phase because the charged
quasiparticle is gapped. We call this phase an excitonic insulator. Then if
the gap is further decreased, the system goes through a second-order
phase transition at some value $U_{\text{ind}}^1<0$ into a phase still
with $\Delta\neq0$, but there are Fermi surfaces in the spectrum of
quasiparticle excitation. Hence this is a conducting phase with
$\Delta\neq0$. A typical image of the Fermi surfaces in the system is
shown in Fig. \ref{fig:exotic-fs}. Eventually, the system goes through
another first order phase transition at
$U_{\text{ind}}^1<U_{\text{ind}}^2$ into a state with $\Delta=0$. This
is a simple semimetal conducting state. A typical phase diagram for
given masses and $g$ is shown in Fig. \ref{fig:a3scan}. For given
masses, the two dimensional phase diagram for different $g$ and
$U_{\text{ind}}$ is shown in Fig. \ref{fig:a3g}.

In phase with the excitonic condensate, 
the ground state has a $U(1)\times U(1)\times
U(1)$ degeneracy because of the corresponding symmetry in the
simplified Hamiltonian. This degeneracy will be lifted to a $U(1)$
degeneracy once we take into account the parts in the interaction
(\ref{eq:coul-gen-hd-q}) which we ignored in the dominant term
approximation. In other words, the full interaction breaks $U(1)\times
U(1)$ symmetry, and only one $U(1)$ is spontaneously broken.

Besides the dominant term in Eq.~(\ref{eq:int-dta-ori}), the next
leading term is the direct interaction in Eq.~(\ref{eq:dir}),
because it is unscreened. That term can be expressed using charge
density operator as
\begin{equation}
  \label{eq:dir-rhoq}
  H_{\text{dir}}=\sum_{q\neq0}\frac{4\pi e^2}{q^2}\left<\rho_q\right>\cdot
  \left<\rho_{-q}\right>
\end{equation}
where $\rho_q$ is the density operator
\begin{equation}
  \label{eq:rhoq}
  \rho_q=\sum_k\left(h_{k+q}^\dagger h_k+d_{k+q}^\dagger d_k\right)
\end{equation}

In a state with translational invariance, the charge density is a
constant, and $\rho_q$ vanishes at any $q\neq0$. However, with an 
excitonic condensate, translational symmetry is actually
broken. This can be understood by the following ways: First, in the
excitonic condensate, there is pairing between holon and three
types of doublon. The total momenta of these pairs are not zero: a
pair of one holon and one doublon $d_\lambda$ has total momentum
$k^D_\lambda$. Therefore the translational symmetry of the system is
broken. Also, one can look at the change of Brillouin zone. Originally
the doublon pockets and holon pocket are at different places in the
Brillouin zone. However, after the Bogoliubov transformation
(\ref{eq:U-k}), the holon and doublon modes are mixed together. The
only way this is allowed is that the Brillouin zone shrinks to one
quarter of its original size, so that $k^D_\lambda$ are all reciprocal
vectors. The shrinking of Brillouin zone means that the unit cell in
real space is enlarged, and the original translational symmetry is
broken.

Because of the broken symmetry, we can look for non-vanishing
expectation value of density operator at $q=k^D_\lambda$. Let us
consider $k^D_1=(\pi, 0)$ as an example. Again we consider only the
excitation near the center of pockets. The expectation value of
density operator is
\begin{equation}
  \label{eq:rhoq-1}
  \left<\rho_{(\pi,0)}\right>=\sum_k\left(
    \left<d_{2k}^\dagger d_{3k}\right>+
    \left<d_{3k}^\dagger d_{2k}\right>
  \right)
\end{equation}

In our ansatz we do not have this average put in explicitly. However,
because the order parameter $\Delta_\lambda$ already breaks the
$U(1)\times U(1)\times U(1)$, we can get a non-vanishing value for
$\left<d_{2k}^\dagger d_{3k}\right>$ from the ansatz.

\begin{figure}[htbp]
  \centering
  \includegraphics{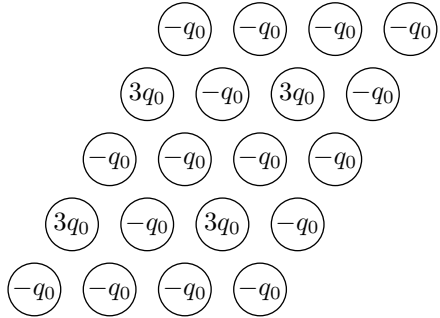}
  \caption{Charge order pattern. $q_0$ can be either positive or
    negative depends on if the relative phase in equation
    (\ref{eq:theta-min}) is $\frac{2\pi}{3}$ or $\frac{5\pi}{3}$.}
  \label{fig:cdw}
\end{figure}

In the mean field state, the average in Eq.~(\ref{eq:rhoq-1}) can
be calculated using the Green's function of doublon. Consider the
Green's function matrix
\begin{equation}
  \label{eq:Gmat}
  G_{hd}(\bm{k},
  \omega)=\left<\Psi_{k}^\dagger(\omega)\Psi_k(\omega)\right>
\end{equation}
Then according to the Hamiltonian in Eq.~(\ref{eq:h-mft-mat}), we
have
\begin{equation}
  \label{eq:Gmat-inv}
  G_{hd}^{-1}(\bm{k}, \omega)=\omega-M_k
\end{equation}
Then the average in Eq.~(\ref{eq:rhoq-1}) can be obtained by
taking the inverse of the above matrix
\begin{equation}
  \label{eq:avg-dd}
  \left<d_{2k}^\dagger d_{3k}\right>
  =\int \frac{d\omega}{2\pi}\frac{\omega-E_{d1}(\bm{k})+\mu}
  {(\omega-E_{k1})(\omega-E_{k2})(\omega-E_{k3})(\omega-E_{k4})}
  \Delta_2^\ast\Delta_3\propto\Delta_2^\ast\Delta_3
\end{equation}
Therefore the average of density operator in equation
(\ref{eq:rhoq-1}) is
\begin{equation}
  \label{eq:rho-1-prop}
  \left<\rho_{(\pi,0)}\right>\propto\re\Delta_2^\ast\Delta_3
  \propto\Delta^2\cos(\theta_2-\theta_3)
\end{equation}
The expectation value of charge density operator at the other two
momenta can be evaluated similarly. Finally we get
\begin{equation}
  \label{eq:Hdir-propto}
  H_{\text{dir}}\propto\frac{4e^2}{\pi}\Delta^2[
  \cos^2(\theta_1-\theta_2)+\cos^2(\theta_2-\theta_3)+
  \cos^2(\theta_3-\theta_1)]
\end{equation}
It can be proved that this energy has a minimum at
\begin{equation}
  \label{eq:theta-min}
  \theta_3-\theta_2=\theta_2-\theta_1=\frac{2\pi}{3}\text{, or }
    \frac{5\pi}{3}
\end{equation}
Such a state breaks the time reversal and parity symmetry because of
the complex phase, but the rotational symmetry of the original lattice
is preserved. Finally from the density operator one can conclude that
there is a charge order in the real space shown in
Fig. \ref{fig:cdw}. From the figure one can see that the unit cell is
a $2\times2$ plaque. This is consistent with the conclusion that the
Brillouin zone shrinks to one quarter of its original size.

\section{Fractionalized metals with $Z_2$ topological order and Luttinger relations}
\label{sec:lutt}

The simplest route from the insulating $Z_2$ spin liquid to a metallic
state is by formation of Fermi surfaces of the holons and doublons. We can easily detect
such a transition as a function of charge density and system parameters by examining
the dispersion relations of the charged excitations in Eq.~(\ref{eq:mix}). These
Fermi surfaces have quasiparticles of charge $\pm e$, but which are spinless.
So the resulting state is not a conventional Fermi liquid, but a $Z_2$ `algebraic charge liquid' \cite{acl}. 
The $Z_2$ topological
order of the insulator survives into the metal, and quasiparticles carry $Z_2$ gauge charge.
This section will address two related questions: what are the Luttinger relations which
constrain the areas enclosed by the Fermi surfaces of this fractionalized metal,
and how does this metal evolve into a Fermi liquid ?

\begin{figure}[htbp]
  \centering
  \includegraphics[width=4in]{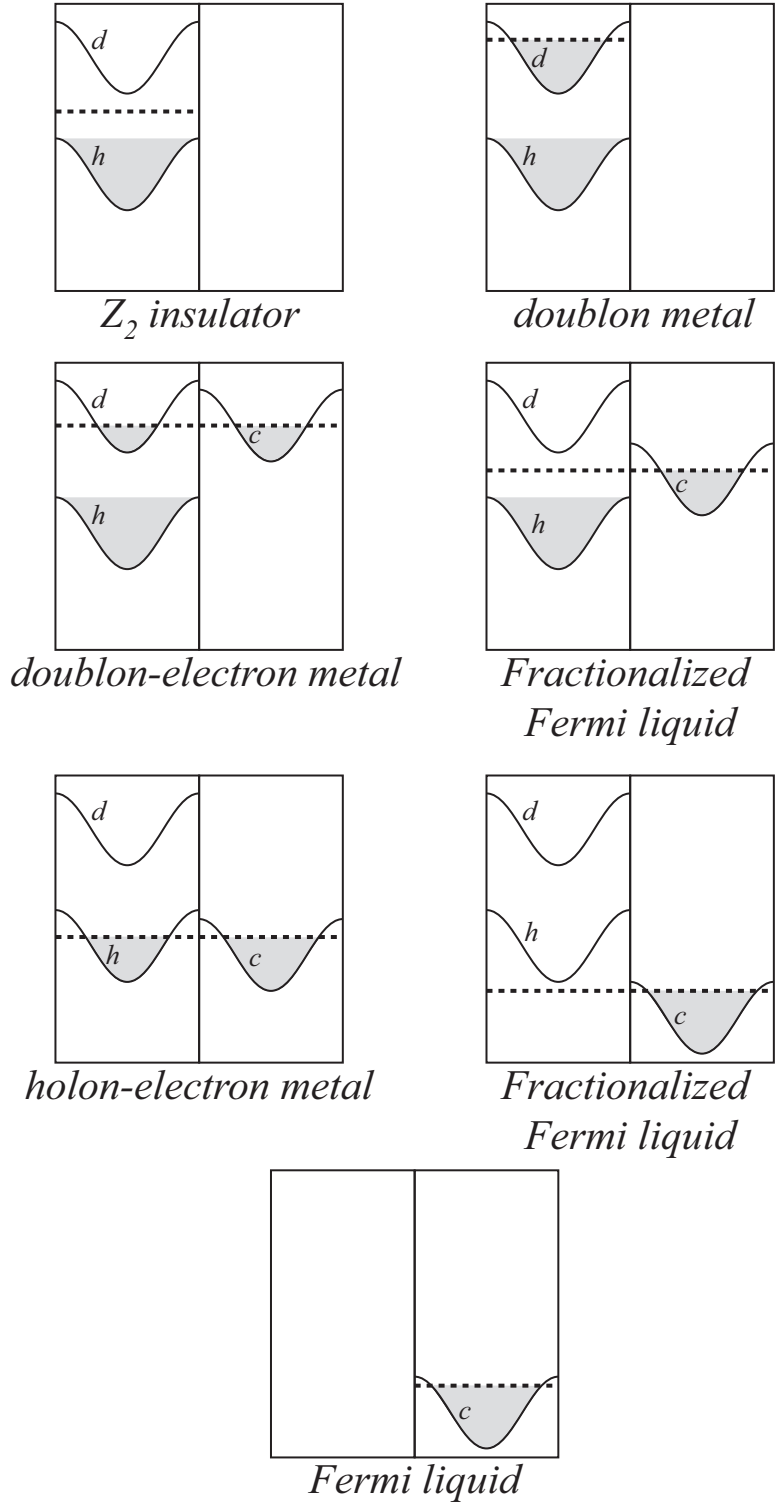}
  \caption{Schematic illustration of the spectrum of fermionic holon ($h$),
  doublon ($d$), and electron ($c_\alpha$) excitations in various
  phases which descend from the $Z_2$ spin liquid. The band structures
  are merely represented by cartoon sketches with the purposes only of 
  indicating the filled states and the location
  of the chemical potential. More realistically, the Fermi surfaces will form around the
  respective dispersion minima in Fig.~\ref{fig:hexagon}.}
  \label{fig:ffl}
\end{figure}

We will examine a route to the Fermi liquid here which is similar to that
considered recently by Kaul {\em et al.} \cite{acl} for related states with U(1)
topological order on the square lattice. We allow for the binding between the holon and
doublon quasiparticles described in Section~\ref{sec:two-band}, and the bosonic
spinons described in Appendix~\ref{sec:lowenergy}. This binding is mediated
not by the gauge force (which is weak in the fractionalized phases we
are discussing here), but by the hopping matrix element $t$ in Eq.~(\ref{eq:tterm}). 
This matrix element is larger than $J$, and so the binding force is strong. 
A computation of the binding force for the low energy holons/doublons and the spinons
is presented in Appendix~\ref{sec:boundstate}, and position of these bound states in the
Brillouin zone is illustrated in Fig.~\ref{fig:hexagon}. 
The resulting bound states will have the quantum numbers of the electron operator in Eq.~(\ref{eq:sc-sep}). Fermi surfaces of 
these `molecular' bound states can also form \cite{powell}, with electron-like quasiparticles which carry charge $\pm e$ and spin 1/2.
Indeed it is possible that these bound states have a lower energy than the holons or doublons, in which case
the first metallic state state will not be an algebraic charge liquid (with holon/doublon Fermi surfaces),
but have electron/hole Fermi surfaces leading to a `fractionalized Fermi liquid' \cite{ffl1,ffl2} (see below
and Fig.~\ref{fig:ffl}). 

So as in Ref.~\onlinecite{acl}, the fractionalized metallic states being considered
here can have a variety of distinct Fermi surfaces. As in a usual Fermi liquid, each Fermi 
surface can be considered to be a surface of either charge $e$ or charge $-e$ 
quasiparticles, with the two perspectives related by a particle-hole transformation. 
There is also a corresponding change in the region of the Brillouin zone which is considered
to be `inside' the Fermi surface, and the value of the area enclosed by the Fermi surface.
For convenience, let us use a perspective in which all Fermi surfaces are described
in terms of charge $-e$ quasiparticles, the same charge as an electron. Thus in the notation
of Eqs.~(\ref{eq:h_k}) and (\ref{eq:g_k_123}), we are considering Fermi 
surfaces of the $h$, $d$, and $c_\alpha$ fermions. Let us by $\mathcal{A}_h$, 
$\mathcal{A}_d$, and $\mathcal{A}_c$ the area of the Brillouin zone enclosed
by these Fermi surfaces, divided by a phase space factor of $4 \pi^2$. We are interested
here in the Luttinger relations which are satisfied by these areas.

As has been discussed in Refs.~\onlinecite{powell,coleman,georges}, there is a one Luttinger
relation associated with each conservation law, local or global. We have the conservation
of total electronic charge which we can write as
\begin{equation}
\frac{1}{N_s} \sum_i \left( g^\dagger_i g_i - f_i^\dagger f_i \right) = x
\label{eq:density}
\end{equation}
where $N_s$ is the number of sites, and $x$ is the density of doped electrons away
from half-filling {\em i.e.\/} the total density of electrons in the band is $1+x$.
There is also the local constraint associated with the Schwinger boson
representation in Eq.~(\ref{eq:sc-sep})
\begin{equation}
f^\dagger_i f_i + g^\dagger_i g_i + \sum_{\alpha} b^\dagger_{i\alpha}
b^{\alpha}_i = 1 \label{eq:slave}
\end{equation}
A key observation is that in any state with $Z_2$ topological order, the global
conservation law associated with Eq.~(\ref{eq:slave}) is `broken', due to the 
appearance of the anomalous pairing condensate $\langle \epsilon_{\alpha\beta}
b_i^\alpha b_j^\beta \rangle$. Consequently\cite{powell}, the Luttinger relation derived from 
Eq.~(\ref{eq:slave}) does not apply in any such state.

We can now derive the constraints on the Fermi surface areas placed by the above
relations using an analysis with exactly parallels that presented in Ref.~\onlinecite{powell}. 
We will not present any details here, but state that in the final result we simply
count the contribution of each term in the conservation law by counting the number
of all such particles inside every Fermi surface, whether as bare particles or as components of 
a molecular state. It is important in this counting to also include the number of 
{\em bosons} {\em i.e.\/} the Luttinger relation applies to bosons 
too \cite{powell,coleman,georges}; of course for the boson number counting, there is a contribution
only from the fermionic molecular states. From such an analysis, after
 properly performing the particle-hole transformations associated with the mapping to the charge $-e$ $h$ 
and $d$ particles, we obtain from Eq.~(\ref{eq:density}) the relation
\begin{equation}
2 \mathcal{A}_c + \mathcal{A}_h + \mathcal{A}_d = 1+x
\label{eq:lutt1}
\end{equation}
This relation applies in all non-superconducting phases.
The corresponding relation from Eq.~(\ref{eq:slave}) is 
\begin{equation}
-\mathcal{A}_h + \mathcal{A}_d = 0
\label{eq:lutt2}
\end{equation}
We reiterate that Eq.~(\ref{eq:lutt2}) applies only in a state without $Z_2$ 
topological order. Indeed, a state without $Z_2$ order cannot have Fermi surfaces
of fractionalized excitations, and so the only sensible solutions of Eq.~(\ref{eq:lutt2}),
when it applies, are $\mathcal{A}_h=\mathcal{A}_g=0$ and $\mathcal{A}_h=\mathcal{A}_g=1$.

With these Luttinger relations at hand, we can now describe their implications for
the various phases. These phases are illustrated in Fig.~\ref{fig:ffl}, and described in 
turn below.

\paragraph{$Z_2$ insulator:} This is the state reviewed in Appendix~\ref{sec:mfa}. The chemical potential is within the band gap of the dispersion in Eq.~(\ref{eq:mix}), as shown
in Fig.~\ref{fig:ffl}.
We have $x=0$, $\mathcal{A}_h=1$, $\mathcal{A}_d=0$, and $\mathcal{A}_c=0$,
and so the relation (\ref{eq:lutt1}) is obeyed. This relation also applies
when an excitonic condensate appears: the only change is that there is mixing between the 
$h$ and $d$ bands.

\paragraph{$Z_2$ metals:} A variety of metallic phases can appear, which are all algebraic
charge liquids \cite{acl},
depending upon the presence/absence of $h$, $d$, and $e$ Fermi surfaces, as shown
in Fig.~\ref{fig:ffl}. All these
phases have $Z_2$ topological order, and so only the sum-rule (\ref{eq:lutt1}) constrains
the areas of the Fermi surfaces. As an example, we can dope the $Z_2$ insulator with negative charge carriers ($x>0$). This will lead to a `doublon metal' with $\mathcal{A}_h=1$, $\mathcal{A}_d=x$ and $\mathcal{A}_c =0$. As the energy of the bound
state between a doublon and spinon is lowered (with increasing $t$), a Fermi surface of electrons can appear, leading to a state with $\mathcal{A}_h=1$, $0<\mathcal{A}_d<x$ and $0<\mathcal{A}_c <x/2$. This fractionalized state with both $h$ and $e$ Fermi 
surfaces (a `doublon-electron' metal) is the particle-hole conjugate of the `holon-hole' metal
discussed recently in Ref.~\onlinecite{acl}. The location of the electron Fermi surface
in this metal is computed in Appendix~\ref{sec:boundstate}, and will reside at pockets
around the electron dispersion minimum in Fig.~\ref{fig:hexagon}.

\paragraph{Fractionalized Fermi liquid:} 
With a continuing increase of the doublon-spinon binding energy in the doublon-electron
metal just discussed, there is a transfer of
states from within the doublon Fermi surface to within the electron Fermi surface, until
we reach the limiting case where $\mathcal{A}_d=0$. At the same time we have $\mathcal{A}_h = 1$, and then Eq.~(\ref{eq:lutt2}) requires $\mathcal{A}_e=x/2$.
So the chemical potential is in the gap between the doublon and holon bands, as shown in Fig.~\ref{fig:ffl}. This state is the fractionalized Fermi liquid of Refs.~\onlinecite{ffl1,ffl2}.

\paragraph{Transition to the Fermi liquid}
We now continue decreasing the energy of the electron bound state, in the hopes
of eventually reaching a conventional Fermi liquid. One plausible route is that
illustrated in Fig.~\ref{fig:ffl}. 
As the electron energy is lowered from the fractionalized Fermi 
liquid, the chemical potential 
goes within the lower holon band, and we obtain a fractionalized holon-electron metal.
Note that this state has both an electron-like Fermi surface with charge $-e$ carriers (as measured by the contribution to the Hall conductivity),
and a holon Fermi surface with charge $+e$ carriers. Eventually, the chemical
potential goes below both the holon and doublon bands (see Fig~\ref{fig:ffl}),
and we obtain another fractionalized Fermi liquid with $\mathcal{A}_h=\mathcal{A}_d=0$
and $\mathcal{A}_c=(1+x)/2$. Note that this is the same area of an electron-like
Fermi surface as in a conventional Fermi liquid. However, the $Z_2$ topological order
is still present, which is why we denote it `fractionalized': the fractionalized
holon and spinon excitations are present, but all of them have an energy gap.
Eventually, however, we expect a confinement transition in the $Z_2$ gauge
theory, leading to a conventional Fermi liquid with no fractionalized excitatons. 
Because the Luttinger
relation in (\ref{eq:lutt2}) is saturated by the electron contribution, we expect
this confinement transition will be in the universality class of the `even' $Z_2$
gauge theory, rather than the `odd' $Z_2$ gauge theory \cite{ms,fradkin}.

\section{Superconducting states}
\label{sec:sc}

The projected symmetry group analysis can be extended to analyze the
pairing symmetry in possible superconducting states. In a
superconducting state, the pairing symmetry can be determined by the
relative phases of the expectation value
\begin{equation}
  \label{eq:scord}
  \Delta_{ij}=\left<\epsilon_{\alpha\beta}c_{i\alpha}c_{j\beta}\right>
\end{equation}
at different directions. On the other hand, the PSG determines how the
operator $c_{i\sigma}$ transforms in rotation, and consequently
determines how Eq.~(\ref{eq:scord}) transforms when the bond
$\left<ij\right>$ is rotated to another direction. Therefore the PSG
of the electron operators can determine the pairing symmetry.

In our Schwinger-boson-slave-fermion ansatz, the fermionic holon and
doublon excitations form Fermi seas around the bottom of their
bands. If there are appropriate attractive forces, holons and doublons
will become superconducting, and the pairing of holons or doublons
will make Eq.~(\ref{eq:scord}) nonzero because the spinon is
already paired. Similarly, the pairing symmetry of holon and doublon
can also be determined by their PSGs, which are related to the PSG of
electron according to Eq.~(\ref{eq:sc-sep}). Therefore the
pairing symmetry of electrons and quasiparticles are related.

When determining the PSG of holon and doublon operators, it is assumed
that the physical electron operator $c_{i\sigma}$ is invariant under
symmetry transformations. However, in the superconducting state the
gauge symmetry of electron is broken and the electron operator may
need a gauge transform after a lattice symmetry operation to
recover. The pairing pattern is directly related to the gauge
transform associated with rotation. Assume that the electron operator
transforms in the following way after a $R_{\frac{\pi}{3}}$ operation:
\begin{equation}
  \label{eq:electron-epsg}
  c_{i\sigma}\rightarrow c_{i\sigma}e^{i\alpha}
\end{equation}
Then after the rotation the superconducting order parameter
(\ref{eq:scord}) becomes
\begin{equation}
  \label{eq:scord-epsg}
  \Delta_{ij}\rightarrow \Delta_{ij}e^{2i\alpha}
\end{equation}
This equation shows the pairing type of the superconducting state.

If we want a superconducting state with translational and rotational
symmetries, the order parameter (\ref{eq:scord}) should recover after
acting $R_{\frac{\pi}{3}}$ three times. This requires the phase
$\alpha$ satisfies
\begin{equation}
  \label{eq:alpha}
  (e^{2i\alpha})^3=1
\end{equation}
which implies that $\alpha=0, \frac{\pi}{3}, \frac{2\pi}{3}$. It is
easy to see that these choices correspond to $s$-wave and
$d_{x^2-y^2}\pm id_{xy}$ pairing patterns. (Note, that the $d$-wave
pairing breaks time reversal symmetry).

On the other hand, Eq.~(\ref{eq:electron-epsg}) also determines
the PSG of holons and doublons, and therefore determines their pairing
patterns. For the $d-id$ pairing, the result is in Table~\ref{tab:pairing}:

\begin{table}[htbp]
  \centering
  \begin{tabular}{c|c|c}
    \hline
    Particle & $R_{\frac{\pi}{3}}$ & Pairing symmetry \\
    \hline
    $c_{i\alpha}$ & $e^{-i\frac{\pi}{3}}c_{i\alpha}$ & $d$-wave \\
    $f$ & $e^{i\frac{\pi}{6}}f$ & $p$-wave \\
    $g_1$ & $e^{i\frac{\pi}{6}}g_2$ & $p$-wave \\
    $g_2$ & $e^{i\frac{\pi}{6}}g_3$ & $p$-wave \\
    $g_3$ & $e^{i\frac{\pi}{6}}g_1$ & $p$-wave \\
    \hline
  \end{tabular}
  \caption{PSG and pairing symmetry of charge excitations.}
  \label{tab:pairing}
\end{table}

For the $s$-wave pairing, the result is in Table~\ref{tab:pairings}:

\begin{table}[htbp]
  \centering
  \begin{tabular}{c|c|c}
    \hline
    Particle & $R_{\frac{\pi}{3}}$ & Pairing symmetry \\
    \hline
    $c_{i\alpha}$ & $c_{i\alpha}$ & $s$-wave \\
    $f$ & $if$ & $f$-wave \\
    $g_1$ & $ig_2$ & $f$-wave \\
    $g_2$ & $ig_3$ & $f$-wave \\
    $g_3$ & $ig_1$ & $f$-wave \\
    \hline
  \end{tabular}
  \caption{PSG and pairing symmetry of charge excitations.\label{tab:pairings}}
\end{table}

From Table~\ref{tab:pairing} and \ref{tab:pairings} we can see that
$p$-wave pairing of holon and doublon will result in $d$-wave pairing of
electron; $f$-wave pairing of holon and doublon will result in $s$-wave
pairing of electron. In the rest of this section we will show that the
interaction mediated by spinon excitation leads to a $p$-wave pairing
symmetry for holon and doublon, and results in a $d$-wave pairing for
electron.

In the spin liquid state, the excitation of spin degree of freedom is
described as Schwinger boson. The holon acquires a kinetic energy from
the original hopping term by replacing the product of spinon operators
by the mean field average. Moreover, beyond the mean field theory this
term also describes interaction between holon and spinon
excitations. This type of interaction results in an interaction
between two holons by exchanging a pair of spinons. Detailed
calculation will show that this interaction is attractive and can lead
to superconducting pairing of fermionic holons.

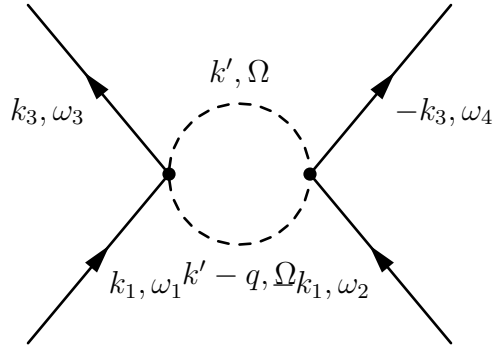
\begin{figure}[htbp]
\centering
\begin{fmffile}{interact-fmf}
\begin{fmfgraph*}(160,160) \fmfpen{thin}
    \fmfbottom{i1,i2}
    \fmftop{o3,o4}
    \fmf{fermion,label=$k_1,,\omega_1$}{i1,v1}
    \fmf{fermion,label=$k_3,,\omega_3$}{v1,o3}
    \fmf{fermion,label=$-k_1,,\omega_2$}{i2,v2}
    \fmf{fermion,label=$-k_3,,\omega_4$}{v2,o4}
    \fmf{dashes,label=$k^\prime,,\Omega$,left}{v1,v2}
    \fmf{dashes,label=$k^\prime-q,,\Omega$,right}{v1,v2}
    \fmfdot{v1,v2}
\end{fmfgraph*}
\end{fmffile}
\caption{\label{fig:inter-feyn}Interaction}
\end{figure}

In Appendix~\ref{sec:interaction}, we calculate the scattering amplitude between two
holons by exchanging two spinons shown in
Fig. \ref{fig:inter-feyn}. Also, a contribution from $O(1/N)$
fluctuation of spinon mean field solution is included to make the
result consistent with the constraint that there is one spinon on
every site.

The scattering between two holons by exchanging two spinons can be
represented as a Feynman diagram in Fig. \ref{fig:inter-feyn}. The
vertex in this figure comes from the hopping term in equation
(\ref{eq:tterm}). For Cooper pairing instability, we are interested in
the following situation: the incoming and outgoing momenta are on the
Fermi surface; the total incoming and outgoing momenta are zero; the
energies of incoming and outgoing particles are on shell. Therefore
the interaction $\Gamma$ is a function of the angle $\theta$ between
incoming and outgoing fermions. In Appendix~\ref{sec:interaction}, we calculate the
scattering amplitude of this Feynman diagram. Also, a contribution
from $O(1/N)$ fluctuation of spinon mean field solution is included to
make the result consistent with the constraint that there is one
spinon on every site. The result is the following

When $m_s\gg q$, we get
\begin{equation}
  \label{eq:gamma-limit1}
  \Gamma\sim -\frac{t^2}{\Lambda^2}\left(m_s+C\frac{q^2}{m_s}\right)
\end{equation}
When $m_s\ll q$, we get
\begin{equation}
  \label{eq:gamma-limit2}
  \Gamma\sim -\frac{t^2}{\Lambda^2}q
\end{equation}
where $\Lambda$ is a UV cutoff of the system, $m_s$ is the spinon gap,
and $q=2k_F\sin\frac{\theta}{2}$ is the momentum transfered in the
interaction.

Since the incoming and outgoing particles are fermions, we need to
antisymmetrized $\Gamma$:
\begin{equation}
  \label{eq:gamma-asym}
  \tilde{\Gamma}(\theta)=\Gamma(\theta)-\Gamma(-\theta)
\end{equation}
The antisymmetrized interaction in the limit of equation
(\ref{eq:gamma-limit1}) is proportional to $\cos\theta$, and therefore
has only $p$-wave component. The antisymmetrized interaction for the
other limit (Eq.~(\ref{eq:gamma-limit2})) is a mixture of
odd-parity wave components, but the $p$-wave component is
dominant. Consequently this interaction will lead to a $p$-wave
superconducting instability for holon, and result in a $d$-wave pairing
superconducting state in terms of electron according to table
\ref{tab:pairing}.

\section{Conclusions}
\label{sec:conc}

This paper has presented a route for the transition from an insulator to a metal for 
electrons on a triangular lattice. Our primary purpose was to pay close attention to the
antiferromagnetic exchange interactions, and the consequent resonating valence bond (RVB)
correlations in the insulator. We were particular interested in the fate of these RVB correlations
in the metal. Our approach borrows from a recent theory \cite{acl} of `algebraic charge liquids' (ACL),
and can be straightforwardly generalized to other lattices in both two and three dimensions.
As we noted in Section~\ref{sec:intro}, our theory of the insulator-to-metal transition
is complementary to the DMFT theory \cite{dmft} of the metal-to-insulator transition.
The latter begins with a theory of the correlated Fermi liquid, but pays scant attention
to RVB correlations, which we believe are particularly important on the triangular lattice.

We have introduced a variety of intermediate phases in the route from the insulator to the metal,
and so we now present a brief review of their basic characteristics. 

First, let us discuss the possible phases of the insulator and their charged excitations. 
We used a $Z_2$ spin 
liquid state \cite{sstri,ashvin} with bosonic spinons as a central point of reference.
This state was reviewed in Appendix~\ref{sec:mfa}, and its low energy excitations are gapped
$S=1/2$ neutral spinons $z_\alpha$ which are located (in our gauge choice) at points in the Brillouin zone shown
in Fig.~\ref{fig:hexagon}. Its charged excitations consist of spinless, charge $+e$ holons $f$, and charge $-e$
doublons $g$, which were described in Section~\ref{sec:holon}; their positions in the Brillouin zone
(in our gauge choice) are also in Fig.~\ref{fig:hexagon}. We also discussed, Appendix~\ref{sec:boundstate} 
the binding of holons/doublons with
spinons to form charged, $S=1/2$ excitations which are hole/electron-like; the location of these bound states
in the Brillouin zone is indicated in Fig~\ref{fig:hexagon}. This binding binding between gapped
excitations is caused by the $t$ 
term in the $t$-$J$ model, and so the attractive force is quite strong. Consequently, it may well be that
the lowest charged excitations of a $Z_2$ spin liquid are electrons and holes, and not doublons and holons.
Note that the binding is present already in
the deconfined spin-liquid state, and is quite distinct from the confinement transition. 

Although, we did not say much about this explicitly, our analysis of excitations in the $Z_2$ spin liquid
is easily continued into confining insulators with N\'eel or VBS order. The transition into the N\'eel state
is driven by the condensation of the $z_\alpha$ spinons, as described in Ref.~\onlinecite{css}. The charged
excitations in the N\'eel state are simple continuations of the holons and doublons in the spin liquid,
as has been discussed in some detail previously for the square lattice \cite{ribhu1}. The charged excitations
in the VBS state will be the electrons and holes described in Appendix~\ref{sec:boundstate}.

Continuing our discussion of insulating states, our primary new result was the prediction of an insulator
with an excitonic condensate in Section~\ref{sec:mft-exciton}. We described the excitonic instability
starting from the $Z_2$ spin liquid, mainly because this led to the simplest technical analysis. However, we
expect very similar excitonic instabilities to appear also from the N\'eel and VBS insulators. 
One physical manifestation of the excitonic instability was a breaking of the space group symmetry
of triangular lattice. In our mean-field analysis we found the translational symmetry breaking depicted
in Fig.~\ref{fig:cdw}, which serves as a test of our theory.

Moving onto metallic states, we found a plethora of possibilities depending upon precisely which
instabilities of the insulator was present (N\'eel, VBS, excitonic), and which of the charged
excitations of the insulator had the lowest energy (holon, doublon, hole, electron). Some of these
states were algebraic charge liquids, while others were Fermi liquids. However, the Fermi surface
areas in all these states satisfied either the conventional Luttinger relation, or a modified Luttinger
relations, and these were discussed in Section~\ref{sec:lutt}.

Finally, we also considered the onset of superconductivity due to pairing of like-charge fermions
in Section~\ref{sec:sc}.
Again, for technical reasons, it was simplest to describe this instability using the holon/doublon
excitations of the $Z_2$ spin liquid, but similar results are expected from instabilities
of the other states.

The possible experimental applications of our results to organic compounds was already discussed
in Section~\ref{sec:intro} and so we will not repeat the discussion here. Here we only reiterate
that our model for $\kappa$-(ET)$_2$Cu$_2$(CN)$_3$ is a $Z_2$ spin liquid, close or across
its transition to a state with N\'eel order, with an excitonic instability at low temperatures.

\acknowledgments

We thank K.~Kanoda, Sung-Sik Lee, and T.~Senthil for useful discussions.
This work was supported by NSF Grant No.\ DMR-0537077. 

\appendix

\section{$Z_2$ spin liquid}
\label{sec:mfa}

Consider the model on a triangle lattice with anti-ferromagnetic
interaction between nearest neighbor sites.
\begin{equation}
  \label{eq:ham-J}
  H=J\sum_{\left<ij\right>}\bm{S}_i\cdot\bm{S}_j
\end{equation}
We use the coordinates system labeled in
Fig.~\ref{fig:coord}. 
and the Schwinger-boson mean field ansatz with the
following ground state expectation values:
\begin{equation}
  \label{eq:ansatz}
  B_{ij}=\left<\sum_\sigma b_{i\sigma}^\dagger
    b_{j\sigma}\right>,\quad
  A_{ij}=\left<\sum_{\alpha\beta}\epsilon_{\alpha\beta}b_{i\alpha}b_{j\beta}\right>
\end{equation}
where $b_{i\sigma}$ is Schwinger boson operator.
\begin{equation}
  \label{eq:spinsum}
  \bm{S}_i=\frac{1}{2}\sum_{\alpha\beta}b_{i\alpha}^\dagger\bm{\sigma}_{\alpha\beta}
  b_{i\beta},
  \quad\sum_\sigma b_{i\sigma}^\dagger b_{i\sigma}=\kappa
\end{equation}

The ansatz (\ref{eq:ansatz}) stays the same if both the spinon
operators and $A$ and $B$ parameters are transformed under the
following gauge transformation
\begin{equation}
  \label{eq:gauge-transf}
  \begin{split}
    b_{i\sigma}\rightarrow e^{i\phi_i}b_{i\sigma} \\
    A_{ij}\rightarrow e^{-i\phi_i-i\phi_j}A_{ij} \\
    B_{ij}\rightarrow e^{i\phi_i-i\phi_j}B_{ij}
  \end{split}
\end{equation}
Because of this gauge freedom, ansatz with different $A_{ij}$ and
$B_{ij}$ configurations may be equivalent after a certain gauge
transform. Therefore the symmetry of $A_{ij}$ and $B_{ij}$ fields may
be smaller than the symmetry of the ansatz. In particular, we consider
the symmetry group of the triangle lattice. (We only consider ansatz
that does not break the lattice symmetry). When a group element $g$
acts on the ansatz, an associated gauge transform $G_g$ may be
required for the $A_{ij}$ and $B_{ij}$ fields to recover. This gauge
transform forms the PSG. According to equation
(\ref{eq:gauge-transf}), gauge transform $G$ can be defined using the
phases $\phi_i(G)$. Consequently the PSG can also be specified using
phases of the gauge transformations. The phases of the gauge
transform associated with symmetry group element $g$ is denoted by
$\phi_g$.

For triangle lattice, the lattice symmetry group can be generated by
four elements: $T_1$ and $T_2$, which are the translations along the 1
and 2 direction; $R_{\frac{\pi}{3}}$, which is the rotation of
$\frac{\pi}{3}$; $\sigma$, which is the reflection operation respect
to the $r_1=r_2$ axis. In the coordinates system used in this note,
these elements can be represented as
\begin{equation}
  \label{eq:transfs}
  \begin{split}
    T_1: (r_1, r_2)\rightarrow(r_1+1, r_2)\\
    T_2: (r_1, r_2)\rightarrow(r_1, r_2+1)\\
    \sigma: (r_1, r_2)\rightarrow(r_2, r_1)\\
    R_{\frac{\pi}{3}}: (r_1, r_2)\rightarrow(r_1-r_2, r_1)\\
  \end{split}
\end{equation}

It can be proved that there are two possibilities for PSG of 2D
triangle lattice. We are interested in one of them, which is called
``zero-flux'' ansatz. The zero-flux PSG can be described as
\begin{equation}
  \label{eq:zero-flux}
  \phi_{T_1}=\phi_{T_2}=\phi_\sigma=0,\quad\phi_{R_{\pi/3}}=\frac{\pi}{2}
\end{equation}
These are also shown in Table~\ref{tab:psg}.

Assume that the system is slightly doped with holes with hole
concentration $\delta\rightarrow0$. Hence the spinon state is not
changed. We describe the system by introducing fermionic holon
\begin{equation}
  \label{eq:spincharge}
  c_{i\sigma}=h_i^\dagger b_{i\sigma}
\end{equation}
where $c_{i\sigma}$ is the electron and $h_i$ is holon operator. To
preserve the physical electron operator, the holon needs to transform
as follows in the gauge transformation (\ref{eq:gauge-transf})
\begin{equation}
  \label{eq:gauge-holon}
  h_i\rightarrow e^{i\phi_i}h_i
\end{equation}
The PSG of holon is summarized in the second row of table
\ref{tab:psg}.

\begin{table}[h]
  \centering
  \begin{tabular}{c|c|c|c|c}
    \hline
    Operator/Field & $T_1$ & $T_2$ & $\sigma$ & $R_{\frac{\pi}{2}}$ \\
    \hline
    $b_{i\sigma}$ & $b_{i\sigma}$ & $b_{i\sigma}$ & $b_{i\sigma}$ & $ib_{i\sigma}$ \\
    $f_i$ & $f_i$ & $f_i$ & $f_i$ & $if_i$ \\
   \hline
  \end{tabular}
  \caption{PSG}
  \label{tab:psg}
\end{table}

\begin{figure}[htbp]
  \centering
  \includegraphics{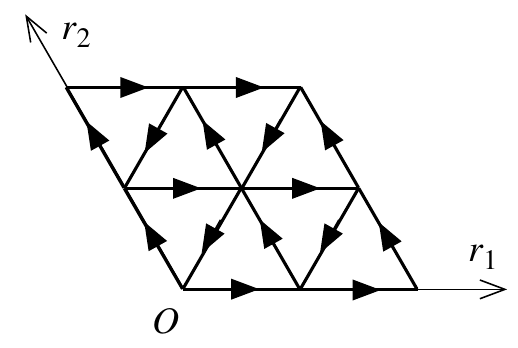}
  \caption{Configuration of $A_{ij}$ in zero-flux
    state\cite{ashvin}. The arrows on the bonds show the direction
    along which $A_{ij}>0$.}
  \label{fig:ansatz}
\end{figure}

The PSG of spinon operators (\ref{eq:zero-flux}) determines how
$A_{ij}$ and $B_{ij}$ transforms under lattice symmetry operations,
and therefore reduces the degree of freedom of parameters in the
ansatz to only two free ones. $\phi_{t_1}=\phi_{T_2}=0$ implies that
the ansatz is translational invariant. By definition $A_{ij}$ and
$B_{ij}$ has the following properties
\begin{equation}
  \label{eq:prop}
  A_{ij}=-A_{ji},\quad B_{ij}=B_{ji}^\ast
\end{equation}
Therefore $\phi_\sigma=0$ and $\phi_{R_{\pi/3}}$ determines the
structure of the ansatz: $B_{ij}\equiv B$ is the same real number on
every bond; $A_{ij}=\pm A$, and the sign is shown in
Fig. \ref{fig:ansatz}.

\subsection{Low energy theory of spinons}
\label{sec:lowenergy}

In this section we analyze low energy excitations of spinon.

For spinon Hamiltonian, if we take the mean field decomposition of
Eq.~(\ref{eq:ham-J}) using parameters (\ref{eq:ansatz}), the
result is
\begin{equation}
  \label{eq:ham-mft}
  \begin{split}
  H_{MF}=&
  J\sum_{\left<ij\right>}\left(-A_{ij}^\ast\hat{A}_{ij}
    +B_{ij}^\ast\hat{B}_{ij}+\text{h. c.}\right)\\
  &+J\sum_{\left<ij\right>}\left(A_{ij}^\ast A_{ij}
    -B_{ij}^\ast B_{ij}+\text{h. c.}\right)
  -\mu\sum_i\left(\sum_\sigma b_{i\sigma}^\dagger
    b_{i\sigma}-\kappa\right)
  \end{split}
\end{equation}
where
\begin{equation}
  \label{eq:ansatzo}
  \hat{B}_{ij}=\sum_\sigma b_{i\sigma}^\dagger
    b_{j\sigma},\quad
  \hat{A}_{ij}=\sum_{\alpha\beta}\epsilon_{\alpha\beta}b_{i\alpha}b_{j\beta}
\end{equation}
Plug in the ``zero-flux'' type of ansatz and write the Hamiltonian in
momentum space, we get

\begin{equation}
  \label{eq:ansatz-0f}
  H_{MF}=\sum_k \Psi_k^\dagger D(\bm{k}) \Psi_k
  +N_s\left[\mu+\mu\kappa-3J(B^2-A^2)\right]
\end{equation}
where
\begin{equation}
  \label{eq:def-psi}
  \Psi_k=(b_{k\uparrow}, b_{-k\downarrow}^\dagger)^T
\end{equation}
and the matrix
\begin{equation}
  \label{eq:def-D}
  D(\bm{k})=\left(
    \begin{matrix}
      JB\re(\xi_k)-\mu & -iJA\im(\xi_k) \\
      iJA\im(\xi_k) & JB\re(\xi_k)-\mu
    \end{matrix}
  \right)
\end{equation}
where $\xi_k$ was defined in Eq.~(\ref{eq:xik}).

The mean field Hamiltonian (\ref{eq:ansatz-0f}) can be diagonalized
using Bogoliubov transformation
\begin{equation}
  \label{eq:bogov}
  \left(\begin{matrix}
      \gamma_{k\uparrow} \\ \gamma_{-k\downarrow}^\dagger
    \end{matrix}\right)=
  \left(\begin{matrix}
      u_k & iv_k \\
      -iv_k & u_k
    \end{matrix}
  \right)
  \cdot\left(\begin{matrix}
      b_{k\uparrow} \\ b_{-k\downarrow}^\dagger
    \end{matrix}\right)
\end{equation}
After diagonalizing the Hamiltonian (\ref{eq:def-D}) becomes
\begin{equation}
  \label{eq:ham-diag}
  H_{MF}=\sum_k\omega(\bm{k})\left(\gamma_{k\uparrow}^\dagger\gamma_{k\uparrow}
    +\gamma_{k\downarrow}^\dagger\gamma_{k\downarrow}+1\right)
  +N_s\left[\mu+\mu\kappa-3J(B^2-A^2)\right]
\end{equation}
The spectrum of the quasi-particle $\gamma_{k\sigma}$ is
\begin{equation}
  \label{eq:wk}
  \omega(\bm{k})=\sqrt{[JB\re(\xi_k)-\mu]^2-[JA\im(\xi_k)]^2}
\end{equation}

The coefficients in Bogoliubov transformation $u_k$ and $v_k$, which
diagonalize the Hamiltonian, are
\begin{equation}
  \label{eq:ukvk}
  \begin{split}
    u_k=\frac{B\re(\xi_k)-\mu+\omega_k}{\sqrt{2\omega_k}} \\
    v_k=\frac{A\im(\xi_k)}{\sqrt{2(B\re(\xi_k)-\mu+\omega_k)\omega_k}}
  \end{split}
\end{equation}
Then it is easy to see that
\begin{equation}
  \label{eq:ukvkminus}
  u_k=u_{-k},\quad v_k=-v_{-k}
\end{equation}

It can be shown that the above spectrum $\omega(\bm{k})$ has two
minimums at $\bm{k}=\pm\bm{k}_C$, where
$\bm{k}_C=(\frac{2}{3}\pi,\frac{2}{3}\pi)$.

To describe the low energy excitation of spinon degree of freedom, the
quasi-particle $\gamma_{k\sigma}$ can be used. Consequently we are
interested in its PSG. Usually the symmetry transforms are described
in coordinate space rather than momentum space. Therefore consider the
Fourier transformation of $\gamma$ operator at a certain momentum
\begin{equation}
  \label{eq:gammai}
  \gamma_{k\sigma}(\bm{r})=\sum_q\gamma_{k+q,\sigma}e^{i\bm{q}\cdot\bm{r}}
\end{equation}
Since the low energy excitations are near $\pm\bm{k}_C$, two new field
operators are defined as
\begin{equation}
  \label{eq:g1g2}
  \begin{split}
    \gamma_{1\sigma}(\bm{r})&=\sum_q\gamma_{k_C+q\sigma}e^{i\bm{q}\cdot\bm{r}} \\
    \gamma_{2\sigma}(\bm{r})&=\sum_k\gamma_{-k_C+q\sigma}e^{i\bm{q}\cdot\bm{r}}
  \end{split}
\end{equation}

Plug in the definition of $\gamma_{k\sigma}$ (\ref{eq:bogov}), one can
get
\begin{align}
  \nonumber
  \gamma_{1\sigma}(\bm{r})\\
  \nonumber
  &=\sum_q\gamma_{k_C+q\sigma}e^{i\bm{q}\cdot\bm{r}}\\
  \label{eq:g1full}
  &=\sum_q\left(u_{k_C+q}b_{k_C+q,\sigma}+ 
    i\sigma v_{k_C+q}b_{-k_C-q,-\sigma}^\dagger\right)
  e^{i\bm{q}\cdot\bm{r}}
\end{align}
Similarly
\begin{equation}
  \label{eq:g2full}
  \gamma_{2\sigma}(\bm{r})=\sum_q\left(u_{-k_C+q}b_{-k_C+q,\sigma}+ 
    i\sigma v_{-k_C+q}b_{k_C-q,-\sigma}^\dagger\right)
  e^{i\bm{q}\cdot\bm{r}}
\end{equation}

Another way of describing the low energy excitations of spinon field
is to use two complex variables $z_1$ and $z_2$, which can also be
used as order parameters in the spin ordered phase.

To begin with, $z$ fields are introduced by discussing the ordered
phase. The excitation spectrum (\ref{eq:wk}) has two minimums at
$\pm\bm{k}_C$. When $\kappa$ becomes larger and larger the gap at
these two momenta vanishes. Consequently the quasi-particle modes at
these momenta will condense and form long range antiferromagnetic
order.

In the spin ordered state, spinon field condensates in the following
way\cite{ashvin}:
\begin{equation}
  \label{eq:spinon-cond}
  \left(
    \begin{matrix}
      \left<b_{i\uparrow}\right>\\
      \left<b_{i\downarrow}\right>
    \end{matrix}
  \right)
  =\left(
    \begin{matrix}
      iz_1& -iz_2\\
      z_2^\ast& z_1^\ast
    \end{matrix}
  \right)\cdot\left(
    \begin{matrix}
      e^{i\bm{k}_c\cdot\bm{r}_i}\\
      e^{-i\bm{k}_c\cdot\bm{r}_i}
    \end{matrix}
  \right)
\end{equation}
where $z_1$ and $z_2$ are the same fields as $c_1$ and $c_2$ in
Ref.~\onlinecite{ashvin}. In antiferromagnetic ordered phase, they acquires
non-zero expectation values. However, in spin liquid phase they
represent gapped excitations.

From Eq.~(\ref{eq:spinon-cond}) one can solve $z_{1,2}$ in terms
of spinon operators
\begin{equation}
  \label{eq:zs}
  \begin{split}
    z_1=\frac{b_{i\uparrow}+ib_{i\downarrow}^\dagger}{2i}e^{-i\bm{k}_C\cdot\bm{r}}\\
    z_2=\frac{b_{i\uparrow}-ib_{i\downarrow}^\dagger}{-2i}e^{i\bm{k}_C\cdot\bm{r}}
  \end{split}
\end{equation}

Actually the $\gamma$ operators and $z$ fields are correlated to each
other: they become the same thing when the system is close to the
critical point between spin liquid phase and antiferromagnetic long
range order phase. To see this, consider only the long wavelength
component in Eq.~(\ref{eq:g1g2}). When $|q|\ll1$, we can take the
approximation $u_{k_C+q}=u_k$, $v_{k_C+q}=v_k$. Therefore $\gamma_1$
becomes
\begin{align*}
  \gamma_{1\sigma}(\bm{r})
  &=\sum_q\gamma_{k_C+q\sigma}e^{i\bm{q}\cdot\bm{r}}\\
  &=\sum_q\left(u_{k_C+q}b_{k_C+q,\sigma}+ 
    i\sigma v_{k_C+q}b_{-k_C-q,-\sigma}^\dagger\right)
  e^{i\bm{q}\cdot\bm{r}} \\
  &\simeq u_{k_C}\sum_qb_{k_C+q,\sigma}e^{i\bm{q}\cdot\bm{r}}+ 
    i\sigma v_{k_C}\sum_qb_{-k_C-q,-\sigma}^\dagger
  e^{i\bm{q}\cdot\bm{r}}
\end{align*}
Note, that
\[
\sum_qb_{k_C+q,\sigma}e^{i\bm{q}\cdot\bm{r}_i}=
\sum_q\int d^2r^\prime\,b_{\sigma}(\bm{r}^\prime)e^{-i(\bm{k}_C+\bm{q})\cdot\bm{r}^\prime}
  e^{i\bm{q}\cdot\bm{r}}
=b_\sigma(\bm{r})e^{-i\bm{k}_C\cdot\bm{r}}
\]
and similarly
\[\sum_qb_{-k_C-q,-\sigma}^\dagger
  e^{i\bm{q}\cdot\bm{r}}=b_{-\sigma}(\bm{r})^\dagger
  e^{-i\bm{k}_C\cdot\bm{r}}\]
The $\gamma_1$ operator can be simplified as
\begin{equation}
  \label{eq:g1eff}
  \gamma_{1\sigma}\simeq 
  \left(u_{k_C}b_\sigma(\bm{r})
    +i\sigma v_{k_C}b_{-\sigma}^\dagger(\bm{r})\right)
  e^{-i\bm{k}_C\cdot\bm{r}}
\end{equation}
Similarly for $\gamma_{2\sigma}$
\begin{equation}
  \label{eq:g2eff}
  \gamma_{2\sigma}\simeq 
  \left(u_{k_C}b_\sigma(\bm{r})
    -i\sigma v_{k_C}b_{-\sigma}^\dagger(\bm{r})\right)
  e^{i\bm{k}_C\cdot\bm{r}}
\end{equation}
(We used Eq.~(\ref{eq:ukvkminus}).)

When the system is approaching the critical point (or $A$ and
$B$ are much larger than the gap), from Eq.~(\ref{eq:ukvk}) one
can see that
\begin{equation}
  \label{eq:ABgap}
  v_{k_c}/u_{k_C}\rightarrow 1
\end{equation}
Therefore by comparing Eq.~(\ref{eq:g1eff}), (\ref{eq:g2eff}) and
Eq.~(\ref{eq:zs}), the following relations are obtained
\begin{equation}
  \label{eq:gamma-z}
  \begin{split}
    \gamma_{1\uparrow}\sim iz_1\\
    \gamma_{1\downarrow}^\dagger\sim z_2\\
    \gamma_{2\uparrow}\sim-iz_2\\
    \gamma_{2\downarrow}^\dagger\sim z_1
  \end{split}
\end{equation}

\section{Interaction mediated by spinon excitation}
\label{sec:interaction}

In this section we calculated the interaction between holon or doublon
through exchanging two spinon excitations described in section
\ref{sec:sc}. 

First, consider the interaction between two holons. The interaction is
shown in Fig. \ref{fig:inter-feyn}, in which the interaction vertex
between holons and spinons comes from the hopping term in equation
(\ref{eq:tterm}). When the spinon gap is small comparing to the band
width of spinon, the low energy spinon excitation can be described by
the $z_1$ and $z_2$ complex fields defined in equation
(\ref{eq:zs}). In terms of $z$ field, the interaction is
\[
  H_{\text{int}}=-t\sum_{\left<ij\right>}f_if_j^\dagger
  \left(z_1^\ast(i)z_1(j)e^{i\bm{k}_C\cdot(\bm{r}_j-\bm{r}_i)}
    +z_2^\ast(i)z_2(j)e^{-i\bm{k}_C\cdot(\bm{r}_j-\bm{r}_i)}+\text{h. c.}\right)
  +\text{h. c.}
\]
in which the terms involve high momentum transfers like
$z_1^\ast z_2e^{i\bm{k}_C\cdot(\bm{r}_i+\bm{r}_2)}$ are ignored. After
a Fourier transform one get from the above equation
\begin{align}
  \nonumber
  H_{\text{int}}=&-2t\sum_{k_1k_2k_3k_4}\delta_{k_2+k_3-k_1-k_4}
  f_{k_1}f_{k_2}^\dagger z_1^\ast(\bm{k}_3)z_1(\bm{k}_4)
  \re(j\xi_{k_4-k_2}+j^2\xi_{k_3+k_2})\\
  \label{eq:vertex-full}
  &-2t\sum_{k_1k_2k_3k_4}\delta_{k_2+k_3-k_1-k_4}
  f_{k_1}f_{k_2}^\dagger z_2^\ast(\bm{k}_3)z_2(\bm{k}_4)
  \re(j^2\xi_{k_4-k_2}+j\xi_{k_3+k_2})
\end{align}
where $j$ is a complex number $j=e^{i\frac{2\pi}{3}}$.

Here we are interested in the situation that both the spinon gap and
the doping are small. When the spinon gap is small, the interaction is
dominated by the spinon excitations near the bottom of the minimum of
spinon band. When the doping is small, the Fermi surface is also
small, so that the momentum of fermions in the interaction vertex is
also small. When all the momenta in the interaction vertex are small,
the $\xi$ function in the vertex can be expanded as
\begin{equation}
  \label{eq:xi-expansion}
  \xi_k=3-\frac{1}{2}k^2a^2+\cdots
\end{equation}
where $a$ is the lattice constant. The leading order in the expansion
is a constant, which corresponds to a contact interaction.


When the gap is small, the two $z$ fields describes
bosonic spinon excitations in the vicinities of momenta
$\pm\left(\frac{2\pi}{3}, \frac{2\pi}{3}\right)$. The two branches of
excitations has the same dispersion relation
\begin{equation}
  \label{eq:spinon-disp}
  \omega_k^2\simeq m_s^2+c^2k^2
\end{equation}
where $m_s$ is the spinon gap and $c$ is the spin wave velocity. The
propagator of $z_i$ fields can be worked out from the spinon mean
field Hamiltonian (\ref{eq:ansatz-0f}) through the relation between
the two representations. The result shows bosonic excitations with the
above dispersion relation and a spectrum weight determined by the
order parameters of the spinon ansatz.
\begin{equation}
  \label{eq:green-z-Z}
  G_{z_1}(\bm{k},\omega_n)=\frac{Z}{\omega_n^2+m_s^2+c^2k^2}
\end{equation}
where the spectrum weight
\begin{equation}
  \label{eq:Z}
  Z=\frac{-\mu+JB_1\re\xi_{k_C}+JA_1\im\xi_{k_C}}{2}
\end{equation}
Similarly the Green's function of $z_2$ field can be obtained. The
result is the same as Eq.~(\ref{eq:green-z-Z}).

Using the interaction vertex (\ref{eq:vertex-full}) and the Green's
function (\ref{eq:green-z-Z}) for $z_{1.2}$ fields, the Feynman
diagram Fig. \ref{fig:inter-feyn} can be evaluated. To the leading
order in Eq.~(\ref{eq:xi-expansion}), 
\begin{equation}
  \label{eq:gamma-approx}
  \Gamma=-2(6t)^2\frac{1}{\beta}\sum_{\omega_n}\int\frac{d^2k^\prime}{(2\pi)^2}
  \frac{Z}{\omega_n^2+m_s^2+c^2k^{\prime2}}
  \frac{Z}{\omega_n^2+m_s^2+c^2(\bm{k}^\prime-\bm{q})^2}
\end{equation}
The integral and frequency summation can be worked out at $T=0$
as in Ref.~\onlinecite{sachdev1999qpt}
\begin{equation}
  \label{eq:gamma-approx2}
  \Gamma=-72t^2Z^2\frac{1}{4\pi c^2q}\tan^{-1}\frac{cq}{2m_s}
\end{equation}


When the momenta in the interaction are all small, this is the result
in the leading order. However, if the constraint of the complex fields
in Eq.~(\ref{eq:cp-constraint}) is taken into account, the
contact interaction should give no contribution to the holon
scattering process because the spinon fields appears in the
interaction can be simply replaced by the constant 1. However, in the
above equation we do get a contribution to the scattering amplitude
from that contact interaction. The reason is that we use a large $N$
expansion (or mean field approximation) in determining the low energy
excitation of the spinon field. However, in such an expansion the
constraint (\ref{eq:cp-constraint}) is replaced by a global chemical
potential. Therefore to correctly include the effect of the constraint
we have to take into account the $1/N$ order fluctuation of the
chemical potential field.

The original action of the $z$ fields is a non-linear sigma model.
\begin{equation}
  \label{eq:nls}
  \mathcal{L}=\frac{N}{2g}|\partial_\mu z_\sigma|^2
\end{equation}
with the constraint 
\begin{equation}
  \label{eq:cp-constraint}
  |z_1|^2+|z_2|^2=1
\end{equation}
The partition function of this action is
\begin{equation}
  \label{eq:Z-nls}
  Z=\int\mathcal{D}z^\ast\mathcal{D}z
  \delta(|z_\sigma|^2-1)\exp\left\{
    -\frac{N}{2g}\int d^Dx |\partial_\mu z_\sigma|^2
  \right\}
\end{equation}
where in our case $D=2+1$. In the above action, the delta function that enforces the constraint
(\ref{eq:cp-constraint}) can be replaced by a Lagrangian multiplier
field $\lambda$ as follows
\begin{equation}
  \label{eq:Z-nls-lambda}
  Z=\int\mathcal{D}z^\ast\mathcal{D}z\mathcal{D}\lambda
  \exp\left\{
    -\frac{N}{2g}\int d^Dx 
    [|\partial_\mu z_\sigma|^2+i\lambda(|z_\sigma|^2-1)]
  \right\}
\end{equation}
Integrate out the $z_\sigma$ field and take the large $N$ limit, one
can get the saddle point of $\lambda$ field as $i\lambda=m^2$, where the
mass is determined by the following mean field self-consistent
Eq.~(see Eq.~(5.21) in Ref. \cite{sachdev1999qpt})
\begin{equation}
  \label{eq:saddle}
  \int^\Lambda \frac{d^Dk}{(2\pi)^D} \frac{1}{k^2+m^2}=\frac{1}{g}
\end{equation}
To calculate the correction in the $1/N$ order, the fluctuation of
$\lambda$ field needs to be included. Section 7.2 of
Ref. \cite{sachdev1999qpt} calculated that the propagator of $\lambda$
field is $1/\Pi(q)$, where $\Pi$ is the bare density-density
correlation function of $z$ fields:
\begin{equation}
  \label{eq:pi}
  \Pi(q)=\int
  \frac{d^Dk}{(2\pi)^D}\frac{1}{k^2+m^2}\frac{1}{(k+q)^2+m^2}
  =\frac{1}{4\pi\sqrt{q^2}}\tan^{-1}\left(\frac{\sqrt{q^2}}{2m}\right)
\end{equation}

\begin{figure}[htbp]
\centering
\begin{fmffile}{interact-lambda}
\begin{fmfgraph*}(160,160) \fmfpen{thin}
    \fmfbottom{i1,i2}
    \fmftop{o3,o4}
    \fmf{fermion,label=$k_1,,\omega_1$}{i1,v1}
    \fmf{fermion,label=$k_3,,\omega_3$}{o3,v1}
    \fmf{fermion,label=$-k_1,,\omega_2$}{i2,v2}
    \fmf{fermion,label=$k_3,,\omega_4$}{o4,v2}
    \fmf{dashes,left}{v1,vv1}
    \fmf{dashes,right}{v1,vv1}
    \fmf{dbl_dashes}{vv1,vv2}
    \fmf{dashes,left}{vv2,v2}
    \fmf{dashes,right}{vv2,v2}
    \fmfdot{v1,v2}
\end{fmfgraph*}
\end{fmffile}
\caption{\label{fig:inter-lambda}Interaction}
\end{figure}
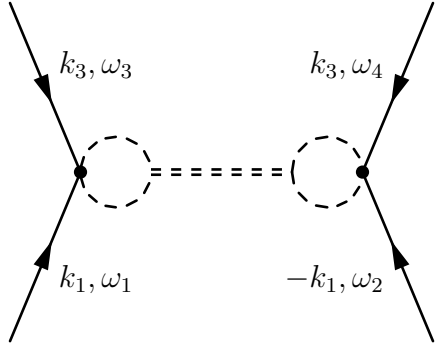

If the fluctuation of $\lambda$ field is included, there will be
another Feynman diagram in addition to Fig. \ref{fig:inter-feyn}
(see Fig. \ref{fig:inter-lambda}).

The contribution from this diagram is
\begin{equation}
  \label{eq:gamma-lambda}
  \Gamma^\prime=(-36t^2)\Pi(q)\cdot(-\Pi(q)^{-1})\Pi(q)=36t^2\Pi(q)
\end{equation}
Therefore it cancels the contribution from
Fig. \ref{fig:inter-feyn}. This is consistent with the constraint
(\ref{eq:cp-constraint}), and it implies that there is no holon-holon
scattering coming from the contact interaction. Consequently we need
to consider interaction with momentum dependence. Taking the next
order in interaction (\ref{eq:vertex-full}), one will get
\begin{align}
  \nonumber
  H_{\text{int}}=&-2t\sum_{k_1k_2k_3k_4}\delta_{k_2+k_3-k_1-k_4}
  f_{k_1}f_{k_2}^\dagger 
  [z_1^\ast(\bm{k}_3)z_1(\bm{k}_4)+z_2^\ast(\bm{k}_3)z_2(\bm{k}_4)]\\
  \label{eq:vertex-k2}
  &\left[6-\frac{1}{2}(\bm{k}_4-\bm{k}_2)^2a^2
    -\frac{1}{2}(\bm{k}_3+\bm{k}_2)^2a^2\right]
\end{align}
where the $a^0$ order gives no contribution to holon-holon interaction
because of the constraint, and therefore can be dropped.

Plug in the interaction in Eq.~(\ref{eq:gamma-lambda}), the
holon-holon interaction shown in Fig. \ref{fig:inter-feyn} and
Fig. \ref{fig:inter-lambda} becomes
\begin{align}
  \nonumber
  \Gamma=&-144t^2\frac{1}{\beta}\sum_{\omega_n}\int
  \frac{d^2k^\prime}{(2\pi)^2}
  \frac{\left[\frac{1}{2}(\bm{k}_1-\bm{k}^\prime)^2a^2+
      \frac{1}{2}(\bm{k}_1+\bm{k}^\prime-\bm{q})^2a^2\right]^2}
  {(\omega_n^2+k^{\prime2}+m_s^2)[\omega_n^2+(\bm{k}^\prime-\bm{q})^2+m_s^2]}\\
  \label{eq:gamma-k2-lambda}
  &+\frac{144t^2}{\Pi(q)}\left\{\frac{1}{\beta}\sum_{\omega_n}\int
  \frac{d^2k^\prime}{(2\pi)^2}
  \frac{\frac{1}{2}(\bm{k}_1-\bm{k}^\prime)^2a^2+
      \frac{1}{2}(\bm{k}_1+\bm{k}^\prime-\bm{q})^2a^2}
  {(\omega_n^2+k^{\prime2}+m_s^2)[\omega_n^2+(\bm{k}^\prime-\bm{q})^2+m_s^2]}
  \right\}^2
\end{align}
Let $\bm{k}_0=\bm{k}_1-\bm{q}/2$, and replace $\bm{k}^\prime$ by
$\bm{k}^\prime+\bm{q}/2$, Eq.~(\ref{eq:gamma-k2-lambda}) can be
simplified as
\begin{align}
  \nonumber
  \Gamma=&-144t^2a^4\frac{1}{\beta}\sum_{\omega_n}\int
  \frac{d^2k^\prime}{(2\pi)^2}
  \frac{\left(k^{\prime2}+k_0^2\right)^2}
  {[\omega_n^2+(\bm{k}^\prime+\bm{q}/2)^2+m_s^2]
    [\omega_n^2+(\bm{k}^\prime-\bm{q}/2)^2+m_s^2]}\\
  \label{eq:gamma-k2-lambda-simp}
  &+\frac{144t^2a^4}{\Pi(q)}\left\{\frac{1}{\beta}\sum_{\omega_n}\int
  \frac{d^2k^\prime}{(2\pi)^2}
  \frac{k^{\prime2}+k_0^2}
  {[\omega_n^2+(\bm{k}^\prime+\bm{q}/2)^2+m_s^2]
    [\omega_n^2+(\bm{k}^\prime-\bm{q}/2)^2+m_s^2]}
  \right\}^2
\end{align}
It is easy to prove that the terms involving $k_0$ are
canceled. Therefore the interaction coefficient can be written as
\begin{align}
  \nonumber
  \Gamma=&-144t^2a^4\frac{1}{\beta}\sum_{\omega_n}\int
  \frac{d^2k^\prime}{(2\pi)^2}
  \frac{k^{\prime4}}
  {[\omega_n^2+(\bm{k}^\prime+\bm{q}/2)^2+m_s^2]
    [\omega_n^2+(\bm{k}^\prime-\bm{q}/2)^2+m_s^2]}\\
  \label{eq:gamma-k2-lambda-final}
  &+\frac{144t^2a^4}{\Pi(q)}\left\{\frac{1}{\beta}\sum_{\omega_n}\int
  \frac{d^2k^\prime}{(2\pi)^2}
  \frac{k^{\prime2}}
  {[\omega_n^2+(\bm{k}^\prime+\bm{q}/2)^2+m_s^2]
    [\omega_n^2+(\bm{k}^\prime-\bm{q}/2)^2+m_s^2]}
  \right\}^2
\end{align}

The $k^{\prime4}$ factor in the integral introduces an UV
divergent. The most divergent term in the integral is proportional to
$\Lambda^3$, where the cut-off $\Lambda=a^{-1}$. However, the most
divergent term does not depend on $q$ and $m$ and will be canceled out
when including the contribution from switching the two outgoing
fermions. Therefore the relevant leading order is $\Lambda^2$. Combine
that with the prefactor $a^4=\Lambda^{-4}$, we conclude that
\begin{equation}
  \label{eq:gamma-scale}
  \Gamma\sim\frac{t^2}{\Lambda^2}
\end{equation}

The dimensionality of $\Gamma$ is $[M^{-1}]$ (see equation
(\ref{eq:gamma-approx2}). Therefore $\Gamma$ should be proportional to
the larger one of $m$ and $q$ when one is much larger than the
other. Consequently by dimensional analysis we can get the results
summarized in Eq.~(\ref{eq:gamma-limit1}) and
(\ref{eq:gamma-limit2}).

Then let us consider the counterpart of this interaction for doublon
excitations. This can be done in the same way as the holon
interactions, while the only difference is that low-energy doublon
excitations are around non-zero momenta $k_\lambda^D$: $(\pi,0)$,
$(0,\pi)$ and $(\pi,\pi)$.

We start with the second term in the hopping Hamiltonian
(\ref{eq:tterm}). In the limit of small doping and small spinon gap,
we can again replace the $b_{i\sigma}$ operators by $z_{1,2}$
fields. After a Fourier transform the interaction vertex becomes
\begin{equation}
  \label{eq:hint-g-ft}
  \begin{split}
    H_{\text{int}}=&2t\sum_{k_1k_2k_3k_4}\delta_{k_2+k_3-k_1-k_4}
    g_{k_1}g_{k_2}^\dagger z_1^\ast(\bm{k}_3)z_1(\bm{k}_4)
    \re(j\xi_{k_2-k_4}+j^2\xi_{k_2+k_3})\\
    +&2t\sum_{k_1k_2k_3k_4}\delta_{k_2+k_3-k_1-k_4}
    g_{k_1}g_{k_2}^\dagger z_2^\ast(\bm{k}_3)z_2(\bm{k}_4)
    \re(j^2\xi_{k_2-k_4}+j\xi_{k_2+k_3})
  \end{split}
\end{equation}

Low energy doublon excitations are located in the vicinities of the
minimums in the doublon band $\bm{k}_{D\lambda}$. To calculate the
interaction between doublons, we need to replace the $g$ operator in
the above Hamiltonian with the operator $g_\lambda$ defined in
Eq.~(\ref{eq:field-doublon}). Then Eq.~(\ref{eq:hint-g-ft})
becomes
\begin{equation}
  \label{eq:hint-g-ft-lambda}
  \begin{split}
    H_{\text{int}}=&2t\sum_{k_1k_2k_3k_4,\lambda}\delta_{k_2+k_3-k_1-k_4}
    g_{\lambda k_1}g_{\lambda k_2}^\dagger z_1^\ast(\bm{k}_3)z_1(\bm{k}_4)
    \re(j\xi_{k_{D\lambda}+k_2-k_4}+j^2\xi_{k_{D\lambda}+k_2+k_3})\\
    +&2t\sum_{k_1k_2k_3k_4}\delta_{k_2+k_3-k_1-k_4}
    g_{\lambda k_1}g_{\lambda k_2}^\dagger z_2^\ast(\bm{k}_3)z_2(\bm{k}_4)
    \re(j^2\xi_{k_{D\lambda}+k_2-k_4}+j\xi_{k_{D\lambda}+k_2+k_3})
  \end{split}  
\end{equation}
Note, that we only included interaction between the same type of
doublon, because interaction between two different types of doublon
cannot be mediated by spinons near the minimum of spinon band, and
therefore is much smaller.

With this interaction vertex and the same Green's function of
$z_{1,2}$ fields we got before in Eq.~(\ref{eq:green-z-Z}), we
can evaluate the Feynman diagram Fig. \ref{fig:inter-feyn} with
incoming and outgoing type-$\lambda$ doublons together with $1/N$
correction in Fig. \ref{fig:inter-lambda}. The result has the same
form as we have in Eq.~(\ref{eq:gamma-limit1}) and
(\ref{eq:gamma-limit2}) because the interaction vertex
(\ref{eq:hint-g-ft-lambda}) has the same form as the one for holons
(\ref{eq:vertex-full}).

\section{Bound state of spinon and holon}
\label{sec:boundstate}

One spinon and one holon may form a bound state if we take into
account the interaction between the two due to the hopping term in
Eq.~(\ref{eq:tterm}). Again, we need to include the effect of
constraint (\ref{eq:cp-constraint}); in other words, we need to
consider the contribution from the fluctuation of $\lambda$ field
shown in Fig. \ref{fig:inter-lambda}.

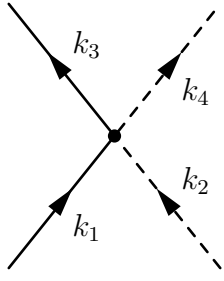
\begin{figure}
\centering
\begin{fmffile}{bound-tree}
\begin{fmfgraph*}(100,100)
	\fmfleft{h1,h2}
	\fmfright{s1,s2}
	\fmf{fermion,label=$k_1$}{h1,v1}
	\fmf{fermion,label=$k_3$}{v1,h2}
	\fmf{scalar,label=$k_4$}{v1,s2}
	\fmf{scalar,label=$k_2$}{s1,v1}
	\fmfdot{v1}
\end{fmfgraph*}
\end{fmffile}
\caption{\label{fig:fbinteract-tree}Interaction between spinon and holon due to 
the hopping term.}
\end{figure}

In general, we consider the interaction between two holons and two spinons with 
incoming and outgoing momenta shown in Fig. \ref{fig:fbinteract-tree}.

The tree level interaction coming from the hopping term in equation
(\ref{eq:vertex-full}) has the momentum dependence according to the
expansion (\ref{eq:xi-expansion})
\begin{equation}
	\label{eq:interac-tree}
	H=t\sum_{k_1k_2k_3k_4}f_{k_2}f_{k_4}^\dagger z_\sigma^\ast(\bm{k}_3)z_\sigma(
	\bm{k}_1)[6-a^2(\bm{k}_1+\bm{k}_2)^2-a^2(\bm{k}_1-\bm{k}_4)^2]
\end{equation}

Therefore the interaction vertex at tree level is
\begin{equation}
	\label{eq:fbgamma-tree}
	\Gamma_{fb}^{(0)}=-t(6-a^2(\bm{k}_1+\bm{k}_2)^2-a^2(\bm{k}_1-\bm{k}_4)^2)
\end{equation}

\begin{figure}[htbp]
\centering
\begin{fmffile}{bound-oneloop}
\begin{fmfgraph*}(300,100)
	\fmfleft{h1,h2}
	\fmfright{s1,s2}
	\fmf{fermion,label=$k_2$}{h1,v1}
	\fmf{fermion,label=$k_4$}{v1,h2}
    \fmf{scalar,left, label=$k_3^\prime$}{v1,v3}
    \fmf{scalar,left, label=$k_1^\prime$}{v3,v1}
    \fmf{dbl_dashes, label=$q$}{v3,v2}
	\fmf{scalar,label=$k_3$}{v2,s2}
	\fmf{scalar,label=$k_1$}{s1,v2}
	\fmfdot{v1}
	\fmfdot{v2}
\end{fmfgraph*}
\end{fmffile}
\caption{\label{fig:fbinteract-oneloop}One loop correction of the
  interaction in Fig. \ref{fig:fbinteract-tree}.}
\end{figure}
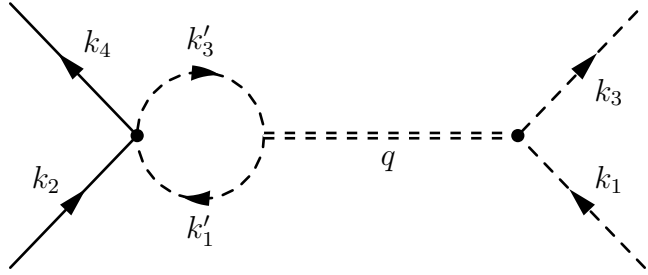

For the same reason we gave in Appendix \ref{sec:interaction}, the one
loop correction due to the fluctuation of $\lambda$ field needs to be
combined with the zero-th order contribution from equation
\ref{eq:interac-tree}. This is represented by
Fig. \ref{fig:fbinteract-oneloop} and is expressed by the following
equation

\begin{equation}
	\label{eq:fbgamma-oneloop}
	\Gamma_{fb}^{(1)}(\bm{k}_1, \bm{q}, i\omega_n)
        =\frac{t}{\Pi(q)}
	\int \frac{d^3k^\prime}{(2\pi)^3}
	\frac{6-a^2(\bm{k}_1^\prime+\bm{k}_2)^2-
          a^2(\bm{k}_1^\prime-\bm{k}_4)^2}
	{[\nu_m^2+(\bm{k}_1^\prime)^2+m_s^2]
    [(\nu_m+\omega_n)^2+(\bm{k}_1^\prime-\bm{q})^2+m_s^2]}
\end{equation}

where the interaction depends on the incoming momentum $\bm{k}_1$,
momentum transfer $\bm{q}$ and energy transfer $\omega_n$.

Add the two contributions in Eq.~(\ref{eq:fbgamma-tree}) and
(\ref{eq:fbgamma-oneloop}) together, one immediately see that the
terms not depend on $\bm{k}_1$ and $\bm{k}_3$ in equation
(\ref{eq:fbgamma-tree}) will be canceled by corresponding terms in the
other term, due to the definition of $\Pi(q)$ in equation
(\ref{eq:pi}). This yields to the following result
\begin{align}
  \nonumber
  &\Gamma_{fb}(\bm{k}_1, \bm{q}, i\omega_n)\\  
  \label{eq:fbgamma-twoparts}
	&=2ta^2\bm{k_1}\cdot\bm{k}_3-
        \frac{ta^2}{\Pi(q)}
        \int \frac{d^3k^\prime}{(2\pi)^3}
	\frac{2\bm{k_1^\prime}\cdot(\bm{k}^\prime_1-\bm{q})}
        {[\nu_m^2+(\bm{k}_1^\prime)^2+m_s^2]
          [(\nu_m+\omega_n)^2+(\bm{k}_1^\prime-\bm{q})^2+m_s^2]}
\end{align}

It is easy to see that in Eq.~(\ref{eq:fbgamma-twoparts}), the
interaction can be seperated into two parts: the first term is a
simple one but with a dependence on $\bm{k}_1$, which does not appear
in usual two-body interactions; the second one is the result of a loop
integral and only depends on the transfered momentum and energy.

The asymptotic behaviors of the second term in $m\ll q$ and $m\gg q$
can be studied by the same dimension analysis method used in Appendix
\ref{sec:interaction}. By counting the power of $k^\prime$ in the loop
integral one can see that the most divergent term is proportional to
$\Lambda$, and the term in the next order does not depend on
$\Lambda$. However, unlike the case in equation
(\ref{eq:gamma-k2-lambda-final}), here the divergent term does give a
$\bm{q}$ and $\omega$ dependent interaction because of the $1/\Pi$
prefactor. Therefore the dominant term in equation
(\ref{eq:fbgamma-twoparts}) is

\begin{equation}
  \label{eq:fbgamma-dominant}
  \Gamma_{fb}\simeq-\frac{ta}{8\pi\Pi(q,\omega)}
  =-\frac{ta c^2\sqrt{c^2q^2-\omega^2}}
  {2\tan^{-1}\frac{\sqrt{c^2q^2-\omega^2}}{2m_s}}
\end{equation}

\begin{figure}[htbp]
  \centering
  \includegraphics{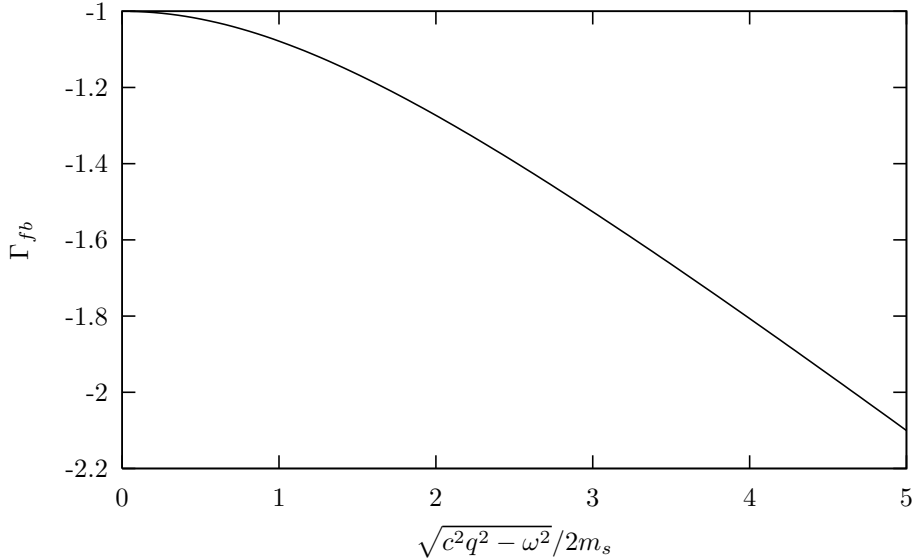}
  \caption{Shape of the interaction function in equation
    (\ref{eq:fbgamma-dominant}).}
  \label{fig:fbgamma}
\end{figure}

And it only depends on $\bm{q}$ and $\omega$. The shape of the
function in Eq.~(\ref{eq:fbgamma-dominant}) is plotted in
Fig. \ref{fig:fbgamma}.

The bound state problem can be solved by solving Bethe-Salpeter
equation\cite{Salpeter1951}. In order to see the qualitative results
of the binding potential in Eq.~(\ref{eq:fbgamma-dominant}) we
ignored the frequency dependence and assume that the dispersion
relations of holon and spinon are both non-relativistic. In this case
the Bethe-Salpeter equation for the relative motion is

\begin{equation}
  \label{eq:BS}
  \left(\frac{p^2}{2\mu}-E_b\right)\phi(\bm{p})
  =-\int \frac{d^2p^\prime}{(2\pi)^2}
  \Gamma(\bm{p}, \bm{p}^\prime, \bm{K})\phi(\bm{p}^\prime)
\end{equation}
where $\bm{p}$ is the relative momentum; $\mu$ is the reduced mass in
center of mass frame; $E_b$ is the bound state energy; $\bm{K}$ is the
total momentum in the center of mass frame; $\phi(\bm{p})$ is the wave
function of relative motion in the momentum space. Numerical
calculation shows that there can be one or more bound states due to
the interaction (\ref{eq:fbgamma-dominant}).

In general, the interaction depends on both the relative momentum
$\bm{p}$, $\bm{p}^\prime$ and the total momentum $\bm{K}$. However,
the dominant term (\ref{eq:fbgamma-dominant}) only depends on
$q=|\bm{p}-\bm{p}^\prime$. Moreover, although the full interaction
(\ref{eq:fbgamma-twoparts}) depends on all momenta, the dependence on
$\bm{K}$ can be separated
\begin{equation}
  \label{eq:fbgamma-separate}
  \Gamma_{fb}=2ta^2K^2+\Gamma_{fb}^\prime(\bm{p}, \bm{p}^\prime)
\end{equation}
Therefore the motion of the center of mass can still be separated from
the relative motion, and the bound state energy can be written as
\begin{equation}
  \label{eq:energy-bound}
  E(\bm{K})=\frac{K^2}{2(m_s+m_h)}+2ta^2K^2+E_b
\end{equation}
where $E_b$ is the bound state energy which does not depend on
$\bm{K}$, and $m_h$, $m_s$ are masses of holon and spinon
respectively. Consequently the bound states have minimal energy at
$\bm{K}=0$. Since the momenta we use in this Appendix are all measured
from the center of spinon and holon energy minimums, the position of
the bound state energy minimum is the sum of holon and spinon
minimums. The former is at the origin; the later is at
$\pm\left(\frac{2\pi}{3}, \frac{2\pi}{3}\right)$. Therefore the
position of bound state minimums is also $\pm\left(\frac{2\pi}{3},
  \frac{2\pi}{3}\right)$. This is shown in Fig. \ref{fig:hexagon}

For doublons, we can also get an attractive interaction between them
and spinon from Eq.~(\ref{eq:hint-g-ft-lambda}). Because of the
same dimension analysis argument the interaction has the same form as
(\ref{eq:fbgamma-dominant}). Therefore, the bound state of doublon and
spinon, or electron excitation locates at the sum of momenta of
doublon and spinon, which is also shown in Fig. \ref{fig:hexagon}.

\end{document}